\documentclass{elsart}
\usepackage{graphicx}
\usepackage{latexsym}
\usepackage{amssymb}
\begin{document}
\newcommand{\YBCO}{ YBa$_{2}$Cu$_{3}$O$_{7}$ }
\newcommand {\LSMO} { La$_{2 / 3}$Sr$_{1 / 3}$MnO$_{3}$ }
\newcommand{\PBCO} { PrBa$_{2}$Cu$_{3}$O$_{7}$ }
\newcommand{\cuo}{CuO$_2$ }
\newcommand{\sto}{SrTiO$_3$ }
\newcommand{\twotheta}{2$\theta$ }
\newcommand{\onebar}{$[1\overline{1}0]$ }
\begin{frontmatter}
\title{Magnetization depinning transition, anisotropic magnetoresistance
and inplane anisotropy in two polytypes of \LSMO epitaxial films.}

\author{Soumen Mandal},
\ead{soumen.mandal@gmail.com}
\author{R C Budhani\corauthref{cor1}}
\address{Condensed Matter - Low Dimensional Systems laboratory,
Department of Physics, Indian Institute of Technology Kanpur, Kanpur - 208016,
India} \ead{rcb@iitk.ac.in} \ead[url]{http://www.iitk.ac.in/cmlds/groupleader.html}
\corauth[cor1]{Corresponding Author. Address: Department of Physics, Indian
Institute of Technology Kanpur, Kanpur - 208016, India, Fax: 91-0512-2590914}

\begin{abstract}
The isothermal magnetoresistance [R($\theta$)] of [001] and [110] epitaxial films of
\LSMO measured as a function of the angle $\theta$ between current ($\vec{I}$) and
magnetic field ($\vec{H}$), both in the plane of the film, is measured at several
temperatures between 10 and 300K. The magnetic easy axis of these polytypes is
intimately related to the orientation of Mn - O - Mn bonds with respect to the
crystallographic axis on the plane of the substrate and energy equivalence of some
of these axes. The magnetization vector ($\vec{M}$) of the [001] and [110] type
films is pinned along the [110] and [001] directions respectively at low fields. A
magnetization orientation phase transition (MRPT) which manifests itself as a
discontinuity and hysteresis in $R(\psi)$ where $\psi$ is the angle between
$\vec{H}$ and the easy axis for the $\vec{H}$ below a critical value $\vec{H}^*$ has
been established. The boundary of the pinned and depinned phase on the H-T plane has
been established. The highly robust pinning of magnetization seen in [110] films is
related to their uniquely defined easy axis. The isothermal resistance R$_\bot$ and
R$_\|$ for $\vec{I} \bot \vec{H}$ and $\vec{I} \| \vec{H}$, respectively for both
polytypes follows the inequality R$_\bot >$ R$_\|$ for all ranges of fields ($0 \leq
H \leq 3500Oe$) and temperatures (10K - 300K). A full fledged analysis of the
rotational magnetoresistance is carried out in the framework of D\"oring theory for
MR in single crystal samples. Strong deviations from the predicted angular
dependence are seen in the irreversible regime of magnetization.
\end{abstract}
\end{frontmatter}

\maketitle \section{Introduction} A galvanomagnetic property of fundamental interest
in thin manganite films is their isothermal magnetoresistance(MR) measured as a
function of the angle($\theta$) between current $\vec{I}$  and applied magnetic
field $\vec{H}$, both in the plane of the film. This angle-dependent
resistivity($\rho(\theta)$) in polycrystalline films of metallic ferromagnets
follows a dependence of the type\cite{Robert,McGuire,Fert,Malozemoff};
\begin{equation} \rho(\theta) = \rho_{\bot} + \left(\rho_{\|} -
\rho_{\bot}\right)\cos^2\theta,\label{eqrho}\end{equation} where $\rho_\|$ and
$\rho_\bot$ are the resistivities for $\vec{I} \| \vec{H}$ and $\vec{I} \bot
\vec{H}$ respectively. The resistivity $\rho(\theta)$, often called the rotational
magnetoresistance (RMR), derives contribution from two carrier scattering processes
one of which depends crucially on the spin - orbit interaction; a
magnetization($\vec{M}$) direction and strength dependent source of anisotropic
scattering. While $\vec{M}$ may not be necessarily collinear with $\vec{H}$ due to
non-zero magnetocrystalline anisotropy, a grain averaging of $\rho$ in a
polycrystalline film yields equation \ref{eqrho}. This contribution to RMR, which
has been known as anisotropic magnetoresistance (AMR), saturates once the field
intensity H exceeds the saturation value H$_s$ beyond which the film becomes a
single domain magnetic entity. Another contribution to RMR comes from the trapping
of mobile carriers in cyclotron orbits due to the Lorentz force. This localizing
effect (a positive contribution to MR), which increases with the carrier mean free
path is called the orbital magnetoresistance (OMR). The magnitude of OMR varies as
square of the magnetic induction $\vec{B}(=\vec{H} + 4\pi\vec{M})$, and shows a
constant positive slope for $H > H_{s}$. While the $\rho_\bot$ is always greater
than $\rho_\|$ for OMR, the relative magnitude $(\rho_\bot/\rho_\|)$ of these
resistivities can be greater or less then unity for AMR. The sign of the inequality
between $\rho_\bot$ and $\rho_\|$ is intimately linked with the electronic band
structure of the material under consideration. In most of the $3d$ - transition
metal alloys $\rho_\| > \rho_\bot$ and the magnetoresistance $(\Delta\rho)/\rho) =
(\rho_\| - \rho_\bot)/\rho_{av}$ can be as large as $\simeq 30\%$ in some dilute Ni
alloys\cite{Robert,McGuire,Fert,Malozemoff}. In low carrier density ferromagnets
such as the hole doped
manganites\cite{Ecksteinapl1996,Ziese,Ziesesing,Infante,Hong}, GaMnAs\cite{Hamaya}
and GdN\cite{Gns} the anisotropic magnetoresistance is small and negative.

The angular dependence of resistivity as expressed by equation \ref{eqrho}, does not
hold in the case of single crystal samples with a non-zero magnetocrystalline
anisotropy. Here the magnetization is not necessarily collinear with $\vec{H}$
because its direction is decided by the competition between the torques exerted on
it by $\vec{H}$ and the crystalline anisotropy field. The $\theta$ dependence of
$\rho$ at low fields, in particular when the torque on $\vec{M}$ is not strong
enough to depin it from the easy axis, may deviate markedly from the behavior
predicted by equation \ref{eqrho}. Often, the $\rho$ vs $\theta$ curve has sharp
jumps and hysteresis suggesting a first order transition. Such magnetization
reorientation phase transition (MRPT) has been seen prominently in Fe film grown on
GaAs\cite{Prinzjap,Prinzprb}.

It is evident that measurements and analysis of the galvanomagnetic property RMR as
a function of field strength and temperature provide rich insight into spin-orbit
interaction, carrier mobilities, scattering length and the torques experienced by
$\vec{M}$ from the applied field $\vec{H}$, and various forms of magnetic
anisotropies which pin $\vec{M}$ along a certain crystallographic direction.
Extensive literature exists on the phenomena of AMR, RMR and OMR on polycrystalline
samples of ferromagnets like Fe, Ni and their
alloys\cite{Robert,McGuire,Fert,Malozemoff}. There are also a large number of
studies on epitaxial thin film of iron in which the deviations from a $\cos^2\theta$
dependence have been addressed in terms of magnetocrystalline
anisotropy\cite{Prinzjap, Prinzprb, Gorkom}.

Extension of the ideas used to understand RMR in elemental ferromagnets to complex
ferromagnetic oxides of  poorly understood magnetocrystalline anisotropy, nature of
charge carriers, their scattering mechanism and their coupling to lattice and spin
degrees of freedom poses major challenges. One family of complex metallic oxides
which has drawn considerable interest in recent years is of manganites of the type
A$_{1-x}$B$_x$MnO$_3$ where A is a rare earth and B an alkaline earth ion.
Electrical conduction in these systems is primarily through $d$-electrons hopping
between neighboring manganese sites\cite{Zener,Anderson,Gennes}. Understanding the
mechanism of RMR in these systems is of fundamental importance.

Most of the studies reported till date on anisotropy of magnetoresistance in
manganites have been carried out on epitaxial films of the average bandwidth
compound La$_{1-x}$Ca$_x$MnO$_3$ (LCMO, x$\approx$0.3) deposited on [100] cut \sto.
These ferromagnetic films have in-plane magnetization  with [001] as the magnetic
easy axis and the magnetic ordering temperature of $\approx$270 K. For example,
measurements of O'Donnell et al\cite{Ecksteinapl1996} show a striking anisotropy and
hysteresis in the low-field MR for $\vec{I}\|\vec{H}$ and $\vec{I}\bot\vec{H}$
configurations, which they attribute to magnetocrystalline anisotropy and colossal
magentoresistance. Ziese and Sena\cite{Ziese}, and Ziese\cite{Ziesesing} have also
measured the low-field resistance anisotropy for $\vec{I}\|\vec{H}$ and
$\vec{I}\bot\vec{H}$ configurations in LCMO films at several temperatures. While the
sign and magnitude of their AMR are consistent with the phenomenological $s-d$
scattering model of Malozemoff\cite{Malozemoff}, its temperature dependence needs
interpretation. Infante et al\cite{Infante} have measured the isothermal MR as a
function of the angle $\theta$ between $\vec{I}$ and $\vec{H}$ at 180 K in [110]
oriented LCMO epitaxial films. They attribute the hysteretic angular dependence of
MR seen at low field to in-plane uniaxial anisotropy of these films. Hong and
coworkers\cite{Hong} have investigated the effect of injected charges, using a field
effect geometry, on AMR of [100] oriented epitaxial La$_{0.7}$Sr$_{0.3}$MnO$_3$
films in order to separate the contributions of carrier concentration and disorder
caused by chemical doping to AMR. While these and other\cite{Tyagi, Snyder} studies
provide a wealth of information on resistance anisotropy in colossal
magnetoresistance manganites, the issues which have remained unaddressed are;
\begin{enumerate}\item Although it is evident from low field measurements of
O'Donnell et al.\cite{Ecksteinapl1996} on [100] LCMO and Infante and
coworkers\cite{Infante} on [110] LSMO films that there is indeed a non-zero in-plane
anisotropy which pins the magnetization vector along the easy axis, the H-T phase
space of the pinned phase is not established in these studies. Moreover, the
relative strength of the in-plane anisotropy, which is directed along different
crystallographic axes for the [100] and [110] films, is not known. There is also a
need to understand the fundamental processes responsible for the anisotropy. \item
At larger fields, the magnetization rotates freely with the field. A systematic
temperature dependence of $\rho(\theta)$ which would allow calculation of
$\Delta\rho(=\rho_\bot - \rho_\|)$ is, however, lacking. A comparative study of
$\Delta\rho$  and its temperature dependence in [100] and [110] films would help in
separating the band structure related contribution and the role of extrinsic effects
such as disorder to anisotropic magnetoresistance. \item While polycrystalline films
of $3d$ ferromagnets show a simple $\cos^2\theta$ dependence of isothermal MR, in
single crystals this is generally not true. Here the orientation of both current and
magnetization with respect to crystal axis is important. In the case of manganites a
full fledged analysis of $\Delta\rho(\theta)$ in terms of D\"oring's
equations\cite{Doring} is lacking. Such a study is desired to establish significant
deviations from the $\cos^2\theta$ dependence of $\Delta\rho$ and the contributions
of other scattering processes to RMR.
\end{enumerate}

Here we present a rigorous study of RMR in two variants of high quality epitaxial
films of \LSMO one with [001] and other with [110] axis normal to the plane of the
substrate. We have analysed the data in the light of D\"oring's equations
\cite{Doring}. The field and temperature dependence of the phenomenological
coefficients between a selected range of temperature has also been reported for both
types of film. We have also drawn the H-T phase diagram for both films to show the
pinned and depinned regions in the H-T phase space. The phase diagram clearly shows
that in case of [110] film the pinning of the magnetization vector is stronger than
that of [001] film.

\section{Experiments}
Thin epitaxial films of \LSMO were deposited on [110] and [001] oriented \sto(STO)
substrates using a multitarget pulsed excimer laser [KrF, $\lambda$ = 248 nm]
ablation technique. The deposition temperature (T$_{d}$), oxygen partial pressure
p$_{O_{2}}$, laser energy density (E$_{d}$) and growth rate (G$_{r}$) used for the
growth of 150 nm thick layers were, 750$^{0}$ C, 0.4 mbar, $\sim$2J/cm$^2$ and
1.3{\AA}/sec respectively. Further details of film deposition are given
elsewhere\cite{Senapati}. The epitaxial growth in two sets of films with [110] and
[001] directions normal to the plane of the film was established with X-ray
diffraction measurements performed in the $\Theta - 2\Theta$ geometry. We have also
examined the surface topography of these epitaxial films using high resolution
scanning electron microscopy (SEM) and atomic force microscopy (AFM) techniques. The
measurements revealed prominent grainyness at nanometer length scale. In the case of
[001] films, the grains were circular of average diameter $\simeq30$nm, while for
the [110] films rectangular grains of area $\simeq$70nm X 40nm were seen. However,
the long edge of the rectangles did not have a preferred orientation. These
isotropic features of the surface texture suggest that the microstructure of these
films cannot contribute to in-plane magnetic anisotropy. For transport measurements,
films were patterned in the form of a $1000 \times 100 {\mu}m^2$ bridge with
photolithography and wet etching such that the long axis of the bridge was parallel
to [001] and [100] direction for the [110] and [001] oriented films respectively.
The measurements of resistivity as a function of temperature, magnetic field
strength and the angle$(\theta)$ between the field and current were performed using
a 4.2K close cycle He - refrigerator with a fully automated home made setup for
applying the field at varying angles between 0 and 2$\pi$ with respect to the
direction of current\cite{Patnaik}. The sample was mounted in a way to keep the
field in the plane of the sample for all values of the angle between $\vec{I}$ and
$\vec{H}$. Isothermal magnetization loops (M-H) were measured for both the samples
using a commercial magnetometer (Quantum Design MPMS XL5 SQUID) by applying the
field at various angles in the plane of the film.

\section{Results and Discussions}
\subsection{Magnetization reorientation phase transition in [110] and [001] films}
\begin{figure}
\centerline{\includegraphics[width=5in, angle = 0]{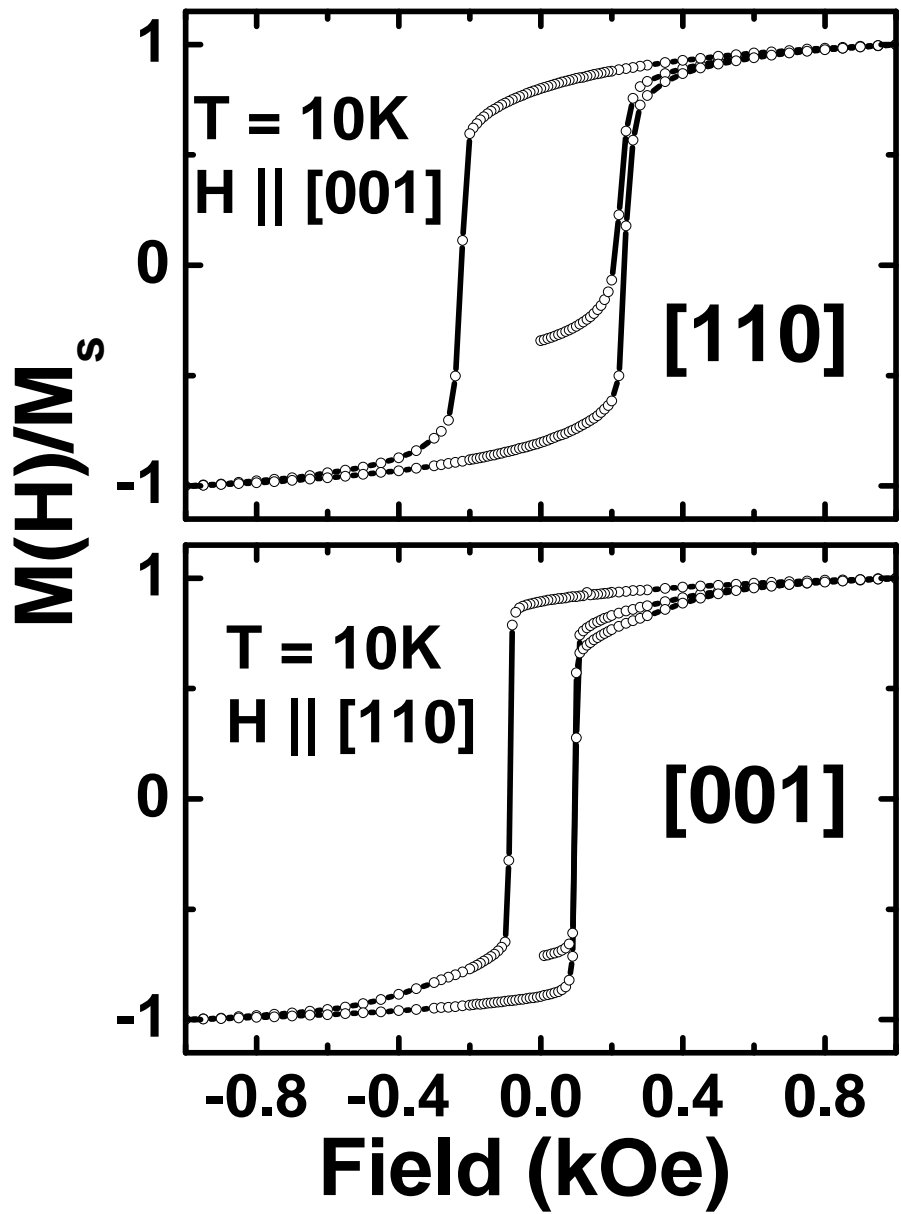}} \caption{Magnetic
hysteresis loops of the [110] and [001] oriented \LSMO films at 10K. The
measurements were done under the zero field cooled condition. The direction of the
applied field in these measurements was along the [001] and [110] crystallographic
directions for the [110] and [001] oriented films respectively. The coercive field
deduced from these measurements is 230 for the [110] and 90 Oe for the [001] samples
respectively.} \label{mhloop}
\end{figure}

Figure \ref{mhloop} shows the magnetization vs field (M-H) loops for the [110] and
[001] epitaxial samples at 10K in terms of the normalized magnetization M(H)/M$_s$,
where M$_s$ is the saturation magnetization. The absolute value of M$_s$ at 10K for
[110] and [100] film is approximately 475 and 400 emu/cm$^3$ respectively. These
numbers have $\approx \pm 30\mbox{emu/cm}^3$ error due to uncertainty in the
measurement of film area and thickness. The magnetic field in these measurements was
applied along the easy axis which is collinear with the [001] and [110]
crystallographic axis for the [110] and [001] epitaxial films
respectively\cite{Easy}. The coercive fields for the [110] and [001] samples deduced
from these measurements are 230 Oe and 90 Oe respectively. The marginally higher
H$_c$ of the [110] film seen here appears to be a common feature of such
films\cite{Suzuki}.

\begin{figure}
\centerline{\includegraphics[width=5in, angle = -90]{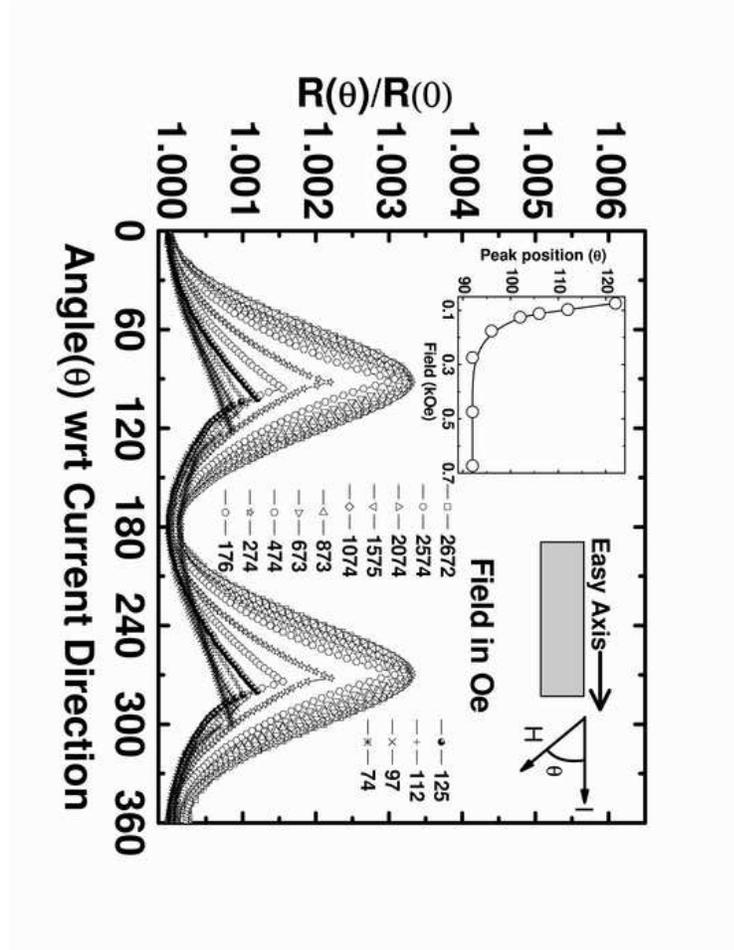}}
\caption{Rotational magnetoresistance R($\theta$) of the [110] film measured at 300K
for different values of the in-plane field. We observe a periodicity of $\pi$ when
the external field H is $\geq$ 300 Oe. Below 300 Oe however, a distinct deviation
from this symmetry is seen, and the peak in resistance is now shifted to $\theta >
90^0$. One noticeable feature of the low field ($<$ 300 Oe) data is a sudden drop in
resistance once the peak value is reached. The top left inset shows the variation of
the position of first peak in the RMR data. A sketch of sample geometry is shown in
the top right hand corner of the figure.}\label{amr110300k}
\end{figure}

\begin{figure} \centerline{\includegraphics[width=5in, angle =
-90]{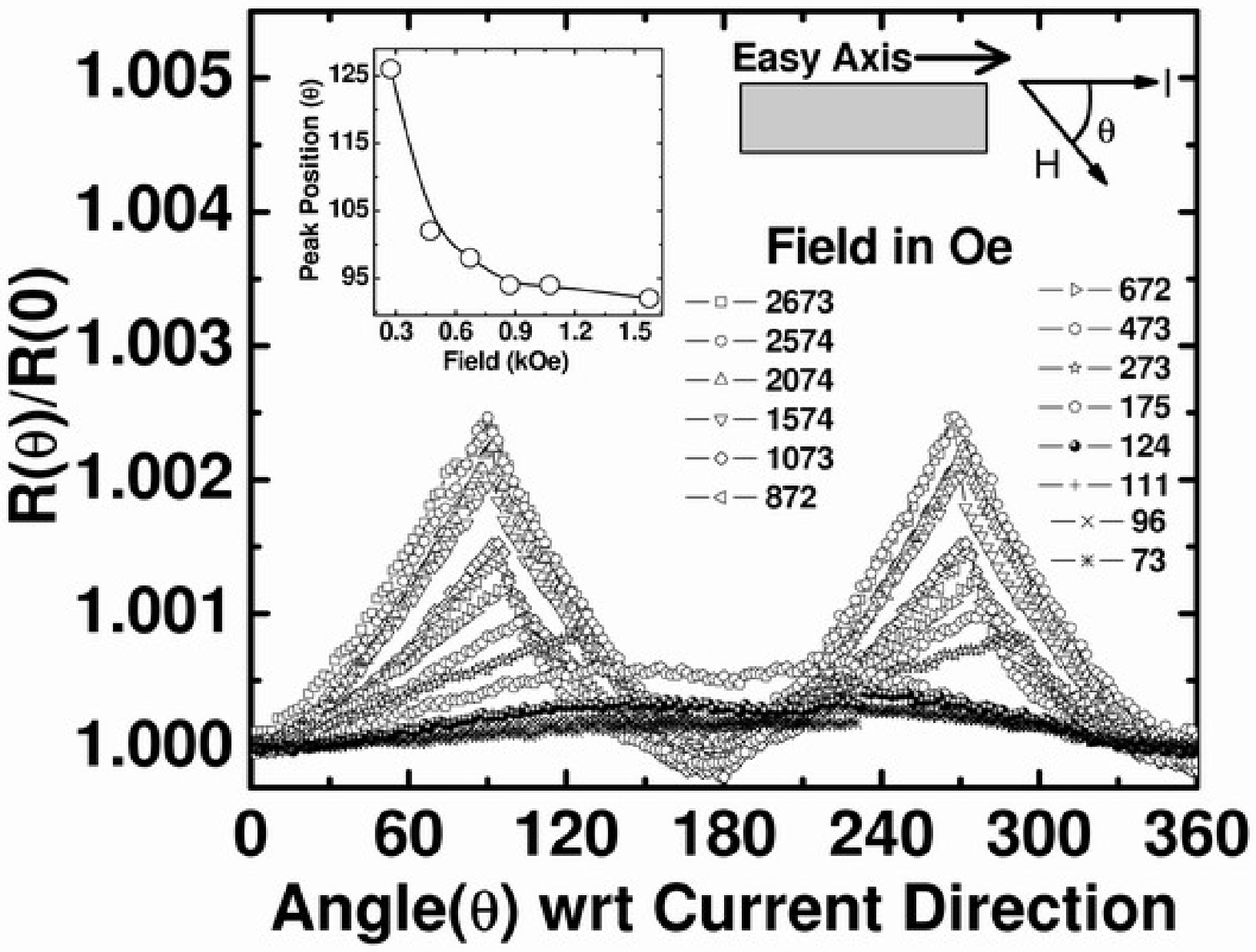}} \caption{The R$(\theta)$/R(0) vs $\theta$ graphs of the [110]
film at 10K for several values of the in-plane field. Here we see deviations from a
symmetric dependence on $\theta$ at fields lower than 1100 Oe. The inset at the top
left hand corner shows the variation of peak position with field of the first peak
in the RMR data. The top right hand corner shows a sketch of the measurement
geometry.} \label{amr11010k}
\end{figure}

The RMR of a [110] oriented LSMO film at 300 and 10K is shown in figure
\ref{amr110300k} and figure \ref{amr11010k} respectively, where we have plotted the
angle ($\theta$) between the directions of current and applied field along the
x-axis and the resistance ratio R($\theta$)/R(0) along the y-axis.

The relevant vectors in the plane of the film are also shown in the right hand inset
of the figures. For the measurements performed at 300K (figure \ref{amr110300k}), we
observe a symmetric R($\theta$)/R(0) curve about $\theta = \pi/2$ and $3\pi/2$ with
a periodicity of $\pi$ when the external field H is $\geq$ 300 Oe. Below 300 Oe
however, there is a distinct deviation from this symmetry; the peak in resistance is
now shifted to $\theta > 90^0$. The variation of the peak position as a function of
field is plotted in the left inset of the figure. One noticeable feature of the low
field ($<$ 300 Oe) data is a sudden drop in resistance once the peak value is
reached. This suggests some kind of a depinning transition. For the R($\theta$)/R(0)
vs $\theta$ curves at 10K (figure \ref{amr11010k}) this deviation from the symmetric
dependence persist upto $\simeq$1100 Oe. Here we also note that at fields below 200
Oe, the resistance has a negligible dependence on the angle between $\vec{I}$ and
$\vec{H}$. One more noteworthy feature of these data is the value of resistance for
$\vec{I} \| \vec{H}$ and $\vec{I} \bot \vec{H}$ configuration. Unlike the case of
3$d$ transition metal films, here $\rho_\bot > \rho_\|$. This is an interesting
feature of the RMR in manganite thin
films\cite{Ecksteinapl1996,Ziese,Ziesesing,Infante,Hong}.

\begin{figure}
\centerline{\includegraphics[width=5in, angle = -90]{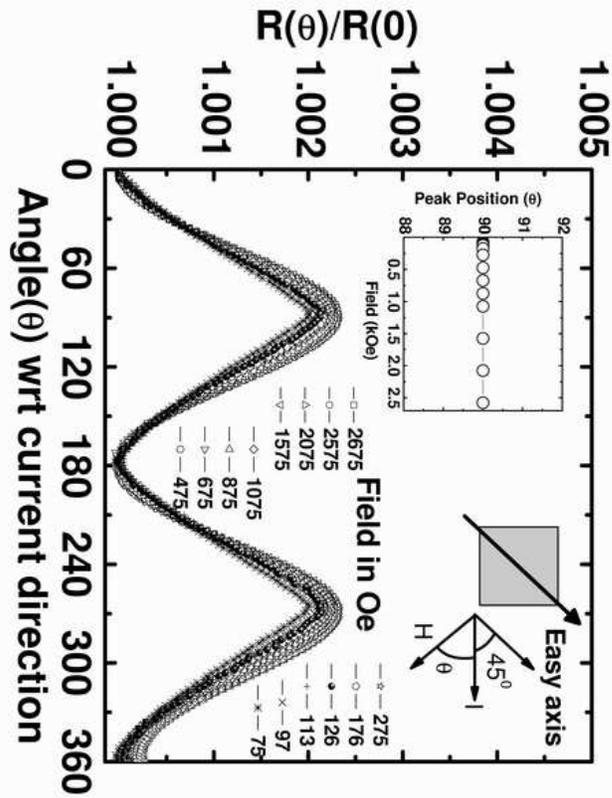}}
\caption{Rotational magnetoresistance R($\theta$) of the [001] LSMO thin film
measured at 300K for different values of the in-plane field. We observe a
periodicity of $\pi$ for all external field H. The top left inset shows the
variation of the position of first peak in the RMR data. A sketch of sample geometry
is shown in the top right hand corner of the figure.} \label{amr100300K}
\end{figure}

\begin{figure}
\centerline{\includegraphics[width=5in, angle = -90]{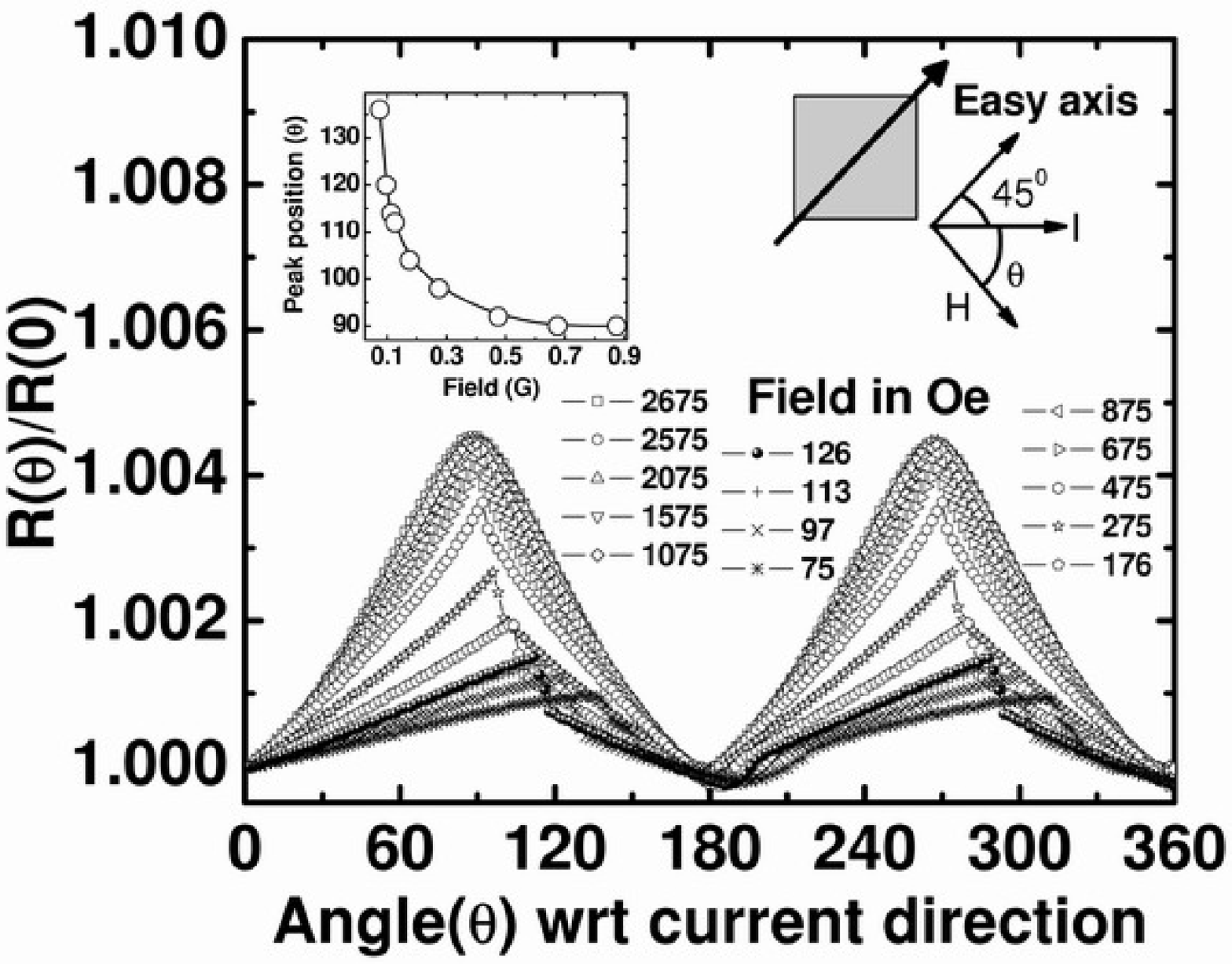}} \caption{The
R$(\theta)$/R(0) vs $\theta$ graphs of the [001] film at 10K for several values of
the in-plane field. Here we see deviations from a symmetric dependence on $\theta$
at fields lower than 500 Oe. The inset at the top left hand corner shows the
variation of peak position with field of the first peak in the RMR data. The top
right hand corner shows a sketch of the measurement geometry.} \label{amr10010K}
\end{figure}

\begin{figure}
\centerline{\includegraphics[width=5in, angle = -90]{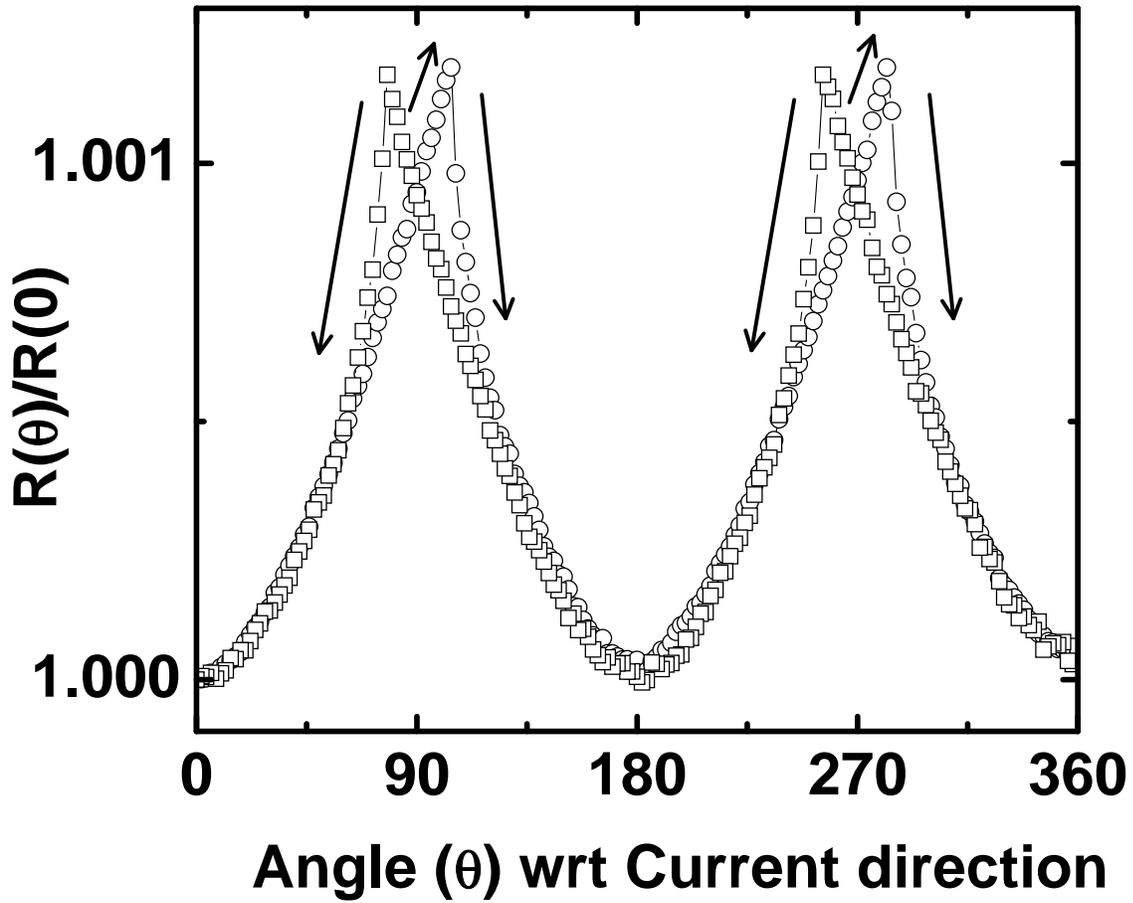}}
\caption{Hysteresis in RMR for [110] LSMO film measured at 300K and 120 Oe . The
open circles and squares show the data when the field is varied in the forward
(0-2$\pi$) and reverse (2$\pi$-0) cycles respectively.} \label{amrhyst}
\end{figure}

\begin{figure}
\centerline{\includegraphics[width=5in, angle = -90]{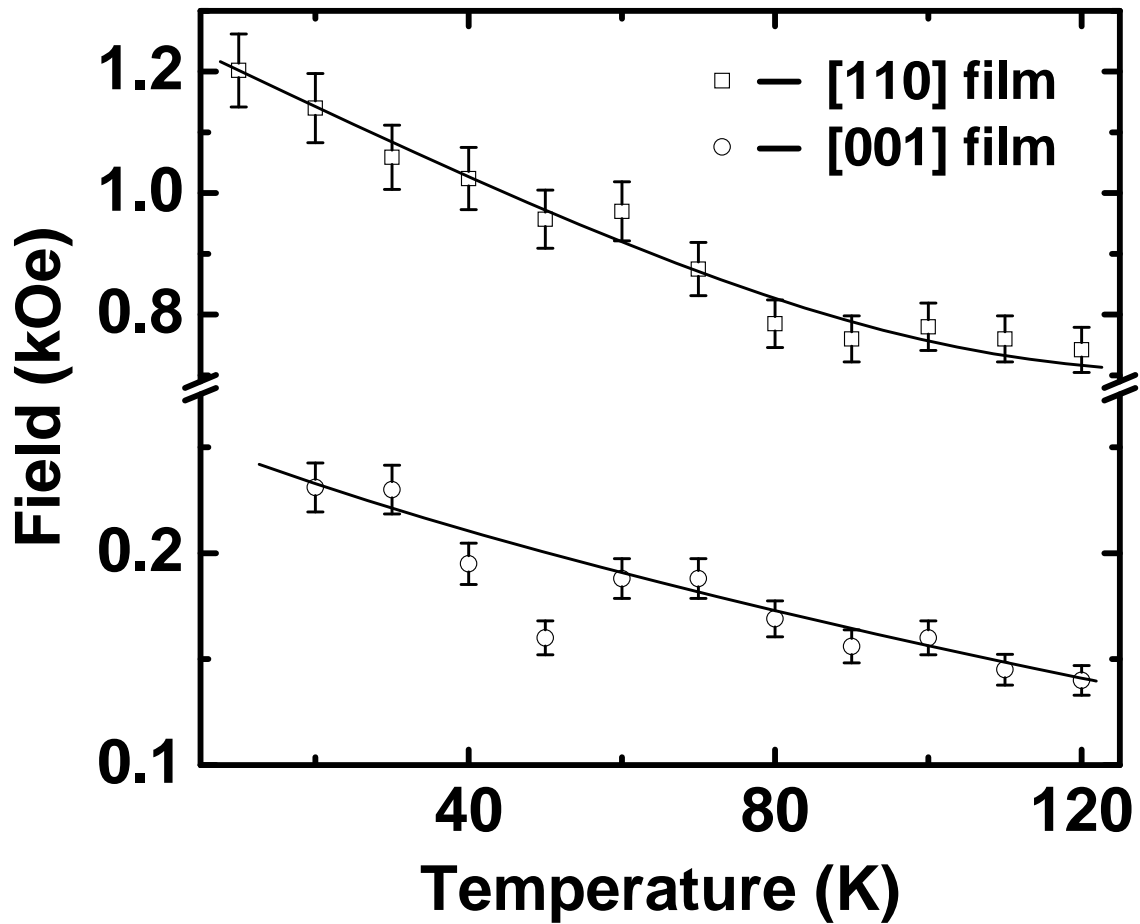}} \caption{H-T
phase diagram for [110](open squares) and [001](open circles) films. The solid lines
are hand drawn to show the most probable separation line between pinned and depinned
states. These data clearly show that the pinning is much stronger in the case of
[110] film than for the [001] films. A change in y-scale emphasizes this point.}
\label{phaseamr}
\end{figure}

In figures \ref{amr100300K} and \ref{amr10010K} we have shown the RMR of the [001]
oriented LSMO film at 300 and 10K respectively. For the measurements at 300K (figure
\ref{amr100300K}) we observe a symmetric angular dependence of the normalized
resistance $R(\theta)/R(0)$ for all values of the applied field, even at fields as
low as 75 Oe.  At 10K however(figure \ref{amr10010K}), the $R(\theta)/R(0)$ deviates
from the symmetric behavior when the field is reduced below $\approx$ 500 Oe. A
sharp drop in resistance when the angle $\theta$ is increased beyond $\theta_{peak}$
for these low field measurements is a remarkable feature of these data. This abrupt
drop in resistivity is accompanied by a hysteresis in the $R(\theta)/R(0)$ vs
$\theta$ plots when the angle is traced back from $2\pi$ to 0. A typical hysteresis
is shown in figure \ref{amrhyst}. The area under the hysteresis loop decreases with
the field. In the left inset of Figs. \ref{amr110300k}, \ref{amr11010k},
\ref{amr100300K} and \ref{amr10010K} we have plotted the $\theta_{peak}$ as a
function of magnetic field strength. As noted from these insets, the peak in RMR
deviates rapidly from $\theta = \pi/2$ as the magnetic field is lowered below a
critical value H*. We have tracked the variation of H* with temperature between 10
and 120K for the two types of films by measuring $\rho(\theta)$ at several fields
while the temperature is held constant. The result of such a measurements is shown
in figure \ref{phaseamr}.

The discontinuous change in $\rho(\theta)$ below a characteristic field H* and
accompanying hysteresis indicate the existence of a magnetization reorientation
phase transition (MRPT)\cite{Prinzjap, Prinzprb, Fisher} in the system driven by the
torque of $\vec{H}$ field on $\vec{M}$. While a rigorous analysis of the MRPT
carried out by minimizing the magnetization free energy functional allows the
calculation of the in-plane magnetocrystalline anisotropies, here we simply argue
that the line in figure \ref{phaseamr} separates the H-T phase space where the
magnetization is pinned along the easy axis and where it is free to rotate with the
field. It is clear from figure \ref{phaseamr} that the magnetization in [110]
oriented film remains pinned along the easy axis over much larger H-T phase space
campared to the [001] oriented film.

\begin{figure} \centerline{\includegraphics[width=4in, angle =
0]{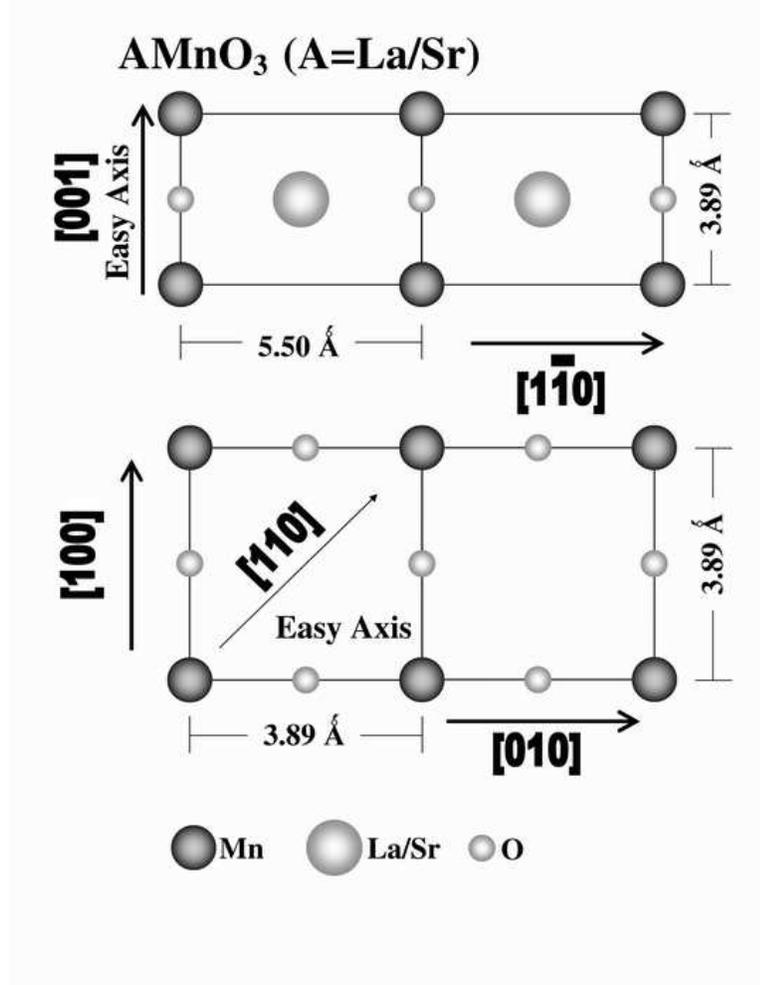}} \caption{Schematic of the ionic positions on the surface of [110]
and [001] films are in the upper and lower panels respectively. For the [110] film,
the two Mn ions along the [001] direction are bridged by an oxygen ion. Hence the
[001] direction acts as magnetic easy axis in these films. In case of [001] the
[100] and [010] direction are degenerate, hence the easy axis is along [110].}
\label{schematic}
\end{figure}

The issue of why the magnetic easy axis in [001] and [110] films of manganites is
different with a different degree of anisotropy energy as suggested by the phase
diagram of figure \ref{phaseamr} has not been addressed in detail although several
workers have reported a difference in in-plane anisotropy axis of [001] and [110]
LSMO films\cite{Suzuki,Tsui,Lecoeur}. It is generally agreed that while for the
[001] films the easy axis of magnetization is along [110] direction, the [110] films
have uniaxial anisotropy with easy and hard directions along [001] and \onebar
respectively. In some earlier studies the magnetic anisotropy in thin LSMO films has
been attributed to epitaxial strain\cite{Suzuki, Tsui}. Berndt etal.\cite{Suzuki}
have shown that for [001] films the anisotropy is dependent on magnetocrystalline
effect while for [110] film it is determined by magnetoelastic effect. Similarly,
Tsui etal.\cite{Tsui} have concluded that the anisotropy of LSMO films is sensitive
to symmetry and morphology of substrate but lattice strains can induce an additional
anisotropy along the direction of tensile strain. Our group has recently
reported\cite{Muduli} dramatic effects of epitaxial strain in magnetic and transport
properties of La$_{1-x}$Sr$_{x}$MnO$_{3}$ (x=0.55) epitaxial film. However, the
effects diminish and bulk-like behavior is seen once the film thickness exceeds
$\simeq100$nm. Since in present case the films are 150nm thick, it is safe to assume
that the strain induced by lattice mismatch between film and substrate is completely
relaxed. Lattice parameter of the films inferred from x-ray diffraction measurement
support the conclusion($d_{[110]} = 3.855${\AA}, $d_{[001]} = 3.85${\AA} and
$d_{bulk} = 3.86${\AA}). Since the topography shows circular and randomly oriented
rectangular grains and x-ray diffraction yields a bulk-like lattice parameter, it
can be concluded safely that the magnetic anisotropy of [001] and [110] films is not
an artifact of strain. However, we expect a fundamental contribution of the
orientation of Mn-O-Mn bonds to magnetic anisotropy. In figure \ref{schematic} we
sketch the atomic arrangement on the top layer of the [001] and [110] cut \sto
crystals and the way epitaxial registry is maintained when \LSMO film grows on the
top. We can see that in the case of [001] oriented film the Mn - O - Mn bonds are
directed along the [100] and [010] direction making them energetically degenerate.
To avoid this degeneracy the magnetization vectors prefer to lie along the [110]
direction making it the easy axis. The difference in the energy of the [110] and
[100]/[010] states of magnetization is expected to be small. This is perhaps the
reason why the depinning field in this case is substantially lower. In the case of
the [110] oriented films, the Mn - O - Mn bond, with a length of $\approx$ 3.89\AA,
is directed along the [001] direction, whereas along the \onebar direction the two
Mn ions are separated by $\approx$ 5.5{\AA } without any bridging oxygen. This makes
the [001] direction the preferred direction for orientation of the magnetization
vector. Furthermore, as the [001] and \onebar directions are highly inequivalent,
the pinning of $\vec{M}$ along [001] is expected to be robust, which is really the
case seen in the phase diagram of figure \ref{phaseamr}.

\subsection{Temperature dependence of RMR}
\begin{figure}
\centerline{\includegraphics[width=5in, angle = -90]{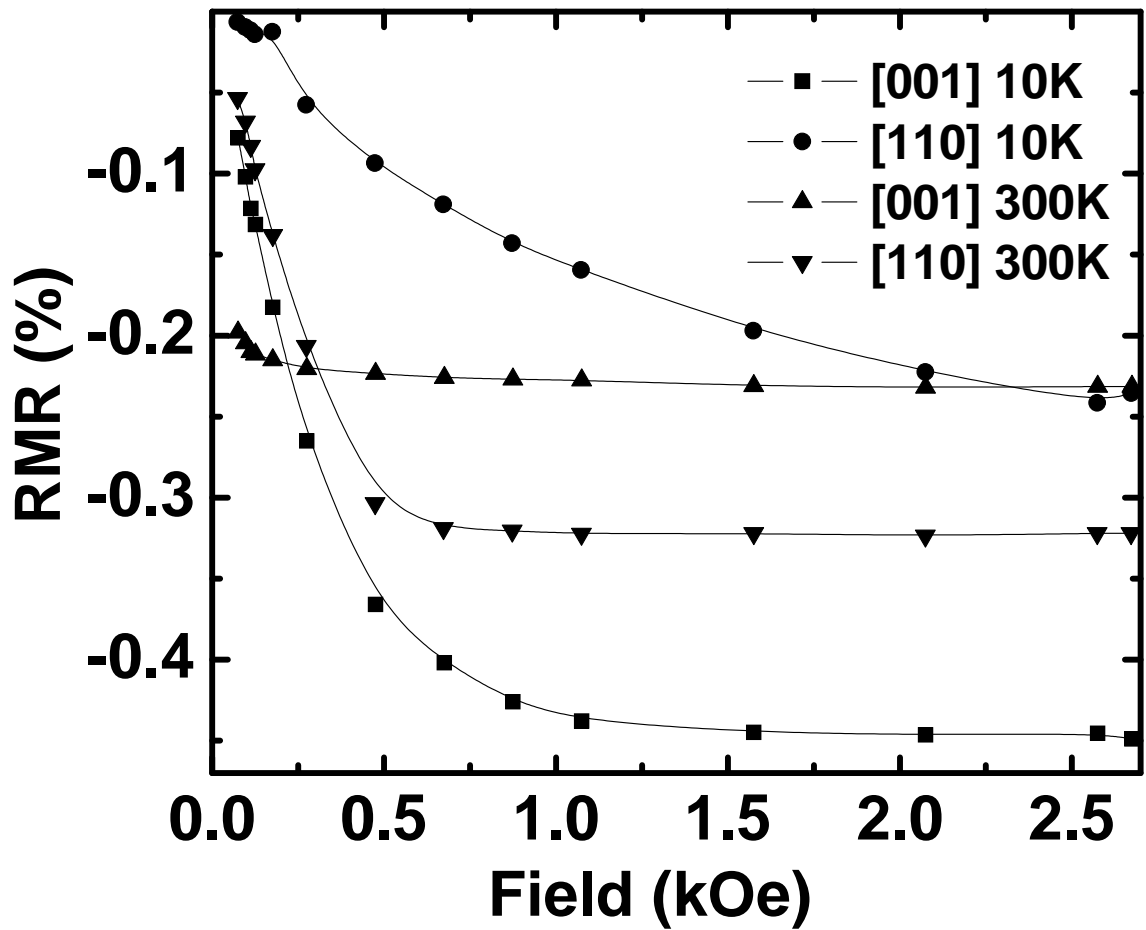}}
\caption{Variation of the RMR percentage, defined as $\frac{R_\| - R_\bot}{R_\bot}
\times 100$, with field for both [110] and [001] LSMO at 10 and 300K. The RMR is
negative for both films.} \label{amrper}
\end{figure}

In figure \ref{amrper} we plot the percentage RMR defined as
$\left[100\left({\rho_\| - \rho_\bot}\right)/\left({\rho_\bot}\right)\right]$ at 10
and 300K for the two types of films as a function of field. The RMR is negative in
both the cases. For the [001] film at 300K, it is also nearly constant at all
fields. For the same film at 10K the magnitude of RMR first increases rapidly with
field and then acquires a saturation value of $\approx -0.46\%$ at H $>$ 1 kOe. For
the [110] film the RMR at 300K saturates to $\sim -0.32\%$ at H $\approx 0.5$ kOe.
The same film has the RMR of $\approx -0.2\%$ at 10K. It is somewhat surprising to
note that the RMR of the [110] film decreases while for [001] it increases as we go
down in temperature.

\begin{figure}
\centerline{\includegraphics[width=5in,angle=-90]{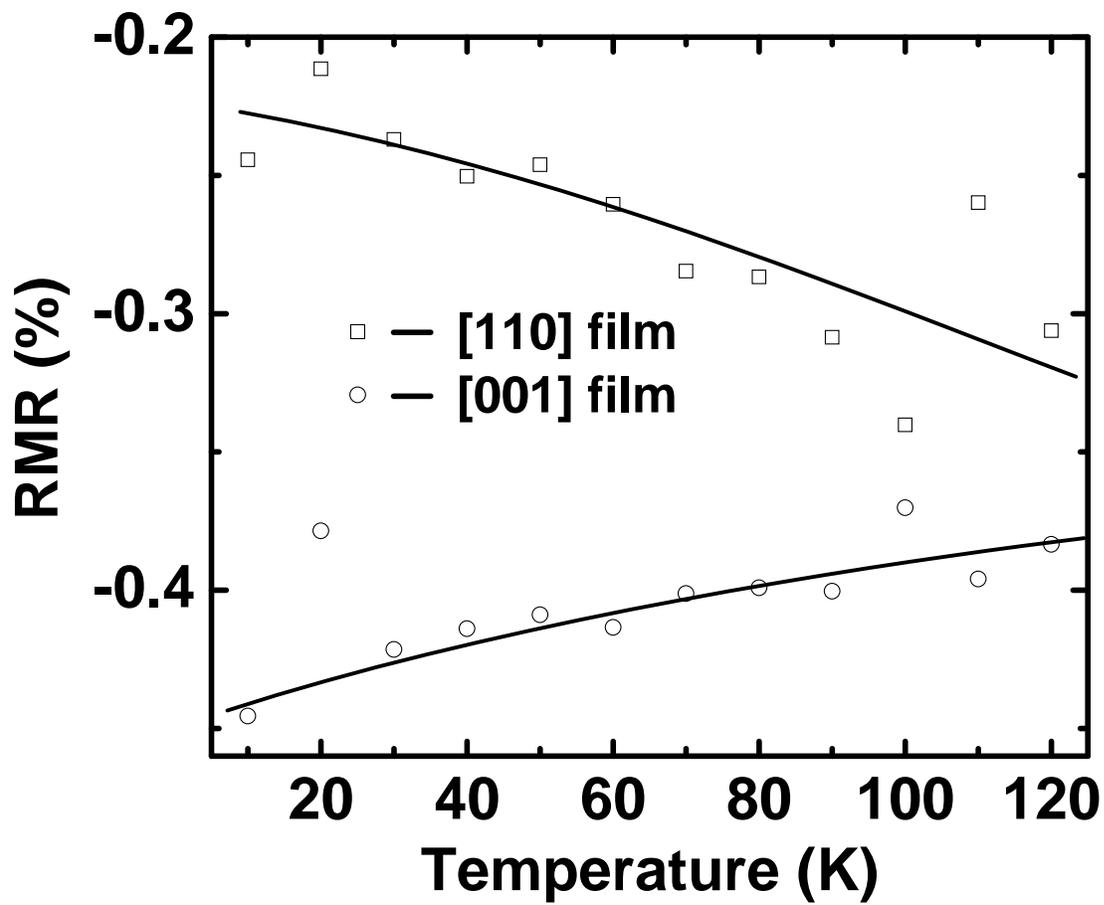}} \caption{Variation
of RMR percentage with temperature for [110](open squares) and [001](open circles)
LSMO films between 10K and 120K. The solid lines are hand drawn depicting the most
probable trend in this temperature range. The field applied in this case 2.5 kOe.}
\label{rmrvartemp}
\end{figure}

In order to address this issue further, we have measured the 2500 Oe RMR of these
films at several temperatures between 10 and 120K. These data are shown in figure
\ref{rmrvartemp}. We note that the RMR of both the samples is negative and its
amplitude in the [001] case decreases monotonically as the temperature is raised to
120K. In fact the RMR deduced from $\rho_\bot(T)$ and $\rho_\|(T)$ measured between
120 and 240K shows that this drop continues till 240K. For the [110] sample however,
the magnitude of RMR first increases with temperature till T $\simeq$ 200K and then
drops on increasing the temperature further.

In the one band model of Malozemoff\cite{Malozemoffprb} as applied to manganites by
Ziese and Sena\cite{Ziese,Ziesesing}, the AMR is given as;
\begin{equation}
\frac{\Delta\rho}{\rho_0} =
-\frac{3}{4}\left[\frac{\lambda^2}{(H_{ex} - \Delta_{cf})^2} -
\frac{\lambda^2}{\Delta_{cf}^2}\right],
\end{equation} where $\lambda$ is the spin - orbit coupling
constant, $\Delta_{cf}$ the crystal field splitting and $H_{ex}$ the exchange field.
By putting in the value of $H_{ex}$, $\Delta_{cf}$ and $\lambda$ for a typical
double exchange manganite they find that $\Delta\rho/\rho_0 \approx -0.85\%$ in the
limit of zero temperature. While a precise temperature dependence of the parameters
$H_{ex}, \Delta_{cf}$ and $\lambda$ is not known, in manganites of $T_c < 300$K a
significant enhancement in AMR near the Curie temperature (T$_c$) has been observed.
Herranz etal.\cite{Herranz,Bibes} have argued that as the Curie temperature is
approached, the double exchange mechanism is impeded by the enhanced Jahn-Teller
distortion of Mn-O octahedron with concomitant unquenching of the orbital angular
momentum which enhances the spin-orbit interaction and hence the AMR. Our data,
however, suggest that in these high quality films of \LSMO where T$_c$ is $\simeq$
360K, the AMR decreases on warming to 300K. While we have not been able to measure
the AMR at T $>$ 300K due to experimental limitations, this is an important issue
that needs to be addressed in future.

\begin{figure}
\centerline{\includegraphics[width=4in,angle=0]{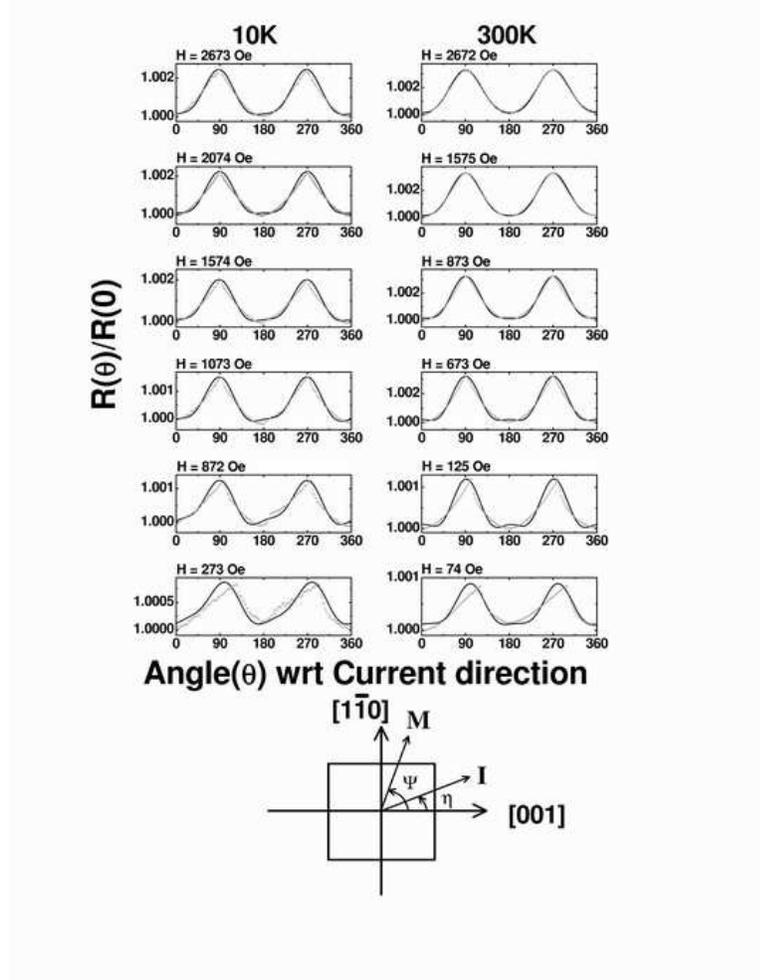}} \caption{Fit of
equation \ref{eqnfin110} to RMR data of [110] film taken at 10 and 300K(left and
right hand side respectively). The dots are actual data and solid lines are fitted
curve. While at 300K a reasonably good fit to equation \ref{eqnfin110} is seen down
to $\simeq$873 Oe, at still lower fields, the angular dependence is characterized by
sharp drop of $R(\theta)/R(0)$ at $\theta > 90^0$. At these fields the torque on
$\vec{M}$ exerted by the external field is not strong enough for coherent rotation.
For the 10K data, the quality of fit is poor even at the higher fields and worsens
as it is reduced below 1574 Oe. A sketch of three vectors $\vec{M}$, $\vec{I}$ and
the unit vector $\hat{n}$, and of relevant angles is shown at the bottom of the
figure. In our case $\vec{I} \| \hat{n}$ hence $\eta = 0$ and [001] is the easy
axis\cite{Easy}.} \label{fit110}
\end{figure}

A rigorous analysis of the RMR data of our samples needs to be done using the
approach of D\"oring\cite{Doring} which entails writing the magnetoresistance of a
cubic ferromagnet as a series in magnetization and current direction cosines
$\alpha_i$ and $\beta_i$ respectively as  \cite{Doring},
\begin{eqnarray} {\frac{\Delta\rho}{\rho_0}} & = &
k_1\left(\alpha_1^2\beta_1^2 +
\alpha_2^2\beta_2^2 + \alpha_3^2\beta_3^2 - \frac{1}{3}\right) \nonumber\\
&& + 2k_2\left(\alpha_1\alpha_2\beta_1\beta_2 +
\alpha_2\alpha_3\beta_2\beta_3 +
\alpha_3\alpha_1\beta_3\beta_1\right) \nonumber\\
&& + k_3(s - c) + k_4\left(\alpha_1^4\beta_1^2 +
\alpha_2^4\beta_2^2 + \alpha_3^4\beta_3^2 + \frac{2}{3}s -
\frac{1}{3}\right)
\nonumber\\
&& + 2k_5\left(\alpha_1\alpha_2\alpha_3^2\beta_1\beta_2 +
\alpha_2\alpha_3\alpha_1^2\beta_2\beta_3 +
\alpha_3\alpha_1\alpha_2^2\beta_3\beta_1\right)\label{doringeq}
\end{eqnarray} where $\rho_0$ is the resistivity at T = 0, $k_i$'s
are phenomenological constants, $c$ is a numerical constant depending on the easy
axis direction, and $s = \alpha_1^2\alpha_2^2 + \alpha_2^2\alpha_3^2 +
\alpha_3^2\alpha_1^2$. For the situation when magnetic domains are distributed
equally among the easy axes in zero applied field. The constant c is 1/4 for [110]
easy axis and zero for [100] easy axis\cite{Ziesesing}.  For the purpose of our
[110] films, the analysis is similar to that used by Gorkom etal \cite{Gorkom} for
[110] Fe with [001] easy axis. Following this work, we can write $\alpha_1 =
-\alpha_2 = \left(1/\sqrt{2}\right)\sin\psi$, $\alpha_3 = \cos\psi$, $\beta_1 =
-\beta_2 = \left(1/\sqrt{2}\right)\sin\eta$ and $\beta_3 = \cos\eta$, where $\psi$
is the angle between the magnetization and the [001] axis and $\eta$, the angle
between the electrical current $\vec{I}$ and the [001] direction. A sketch of three
vectors $\vec{M}$, $\vec{I}$ and the unit vector $\hat{n}$ pointing along the easy
axis in a most generalized situation is given at the bottom of figure \ref{fit110}.
On substituting these parameters in equation \ref{doringeq} one gets,
\begin{eqnarray} \frac{\Delta\rho}{\rho_0} - \delta & = &
C_1\cos^2\psi + C_2\cos^4\psi + C_3\cos\psi\sin\psi +
C_4\cos\psi\sin^3\psi \label{eqnfin110}
\;\;\;\;\;\;\;\;\;\;\;\;\;\;\;\;\;\;\;\;\;\;\;\;\;\;\;\;\;\;
\\ \mbox{where,} && \nonumber\\ C_1 & = & k_1\left(\cos^2\eta -
\frac{1}{2}\sin^2\eta\right) - \frac{k_2}{2}\sin^2\eta +
\frac{k_3}{2} + k_4\left(\frac{1}{3} -
\frac{1}{2}\sin^2\eta\right) \nonumber\\ &&+ \frac{k_5}{2}\sin^2\eta\\
C_2 & = & - \frac{3k_3}{4} + k_4\left(\cos^2\eta +
\frac{1}{4}\sin^2\eta - \frac{1}{2}\right) -
\frac{k_5}{2}\sin^2\eta \\ C_3 & = & 2k_2\cos\eta\sin\eta \\
C_4 & = & k_5\cos\eta\sin\eta \\ \delta & = & k_1\left(\frac{1}{2}\sin^2\eta -
\frac{1}{3}\right) + \frac{k_2}{2}\sin^2\eta + \frac{k_3}{4} +
k_4\left(\frac{1}{4}\sin^2\eta - \frac{1}{6}\right)\end{eqnarray} In our case
$\vec{I} \| \hat{n}$ hence $\eta = 0$. This result leads to $C_1 = k_1 + ({k_3}/{2})
+ ({k_4}/{3}), C_2 = ({-3k_3}/{4}) + ({k_4}/{2}), C_3 = C_4 = 0 \mbox{  and } \delta
= ({-k_1}/{3}) + ({k_3}/{4}) - ({k_4}/{6})$. Since $\delta$ depends only on $k_1,
k_2$ and $k_3$ which in turn are temperature dependent coefficients, we can lump
$\delta$ in $({\Delta\rho})/({\rho_0})$ for an isothermal measurement, and then, the
right hand side of equation \ref{eqnfin110} can be written as $({R - R_0})/({R_0})$,
where $R_0$ is the resistance at the peak position.

\begin{figure}
\centerline{\includegraphics[width=5in,angle=-90]{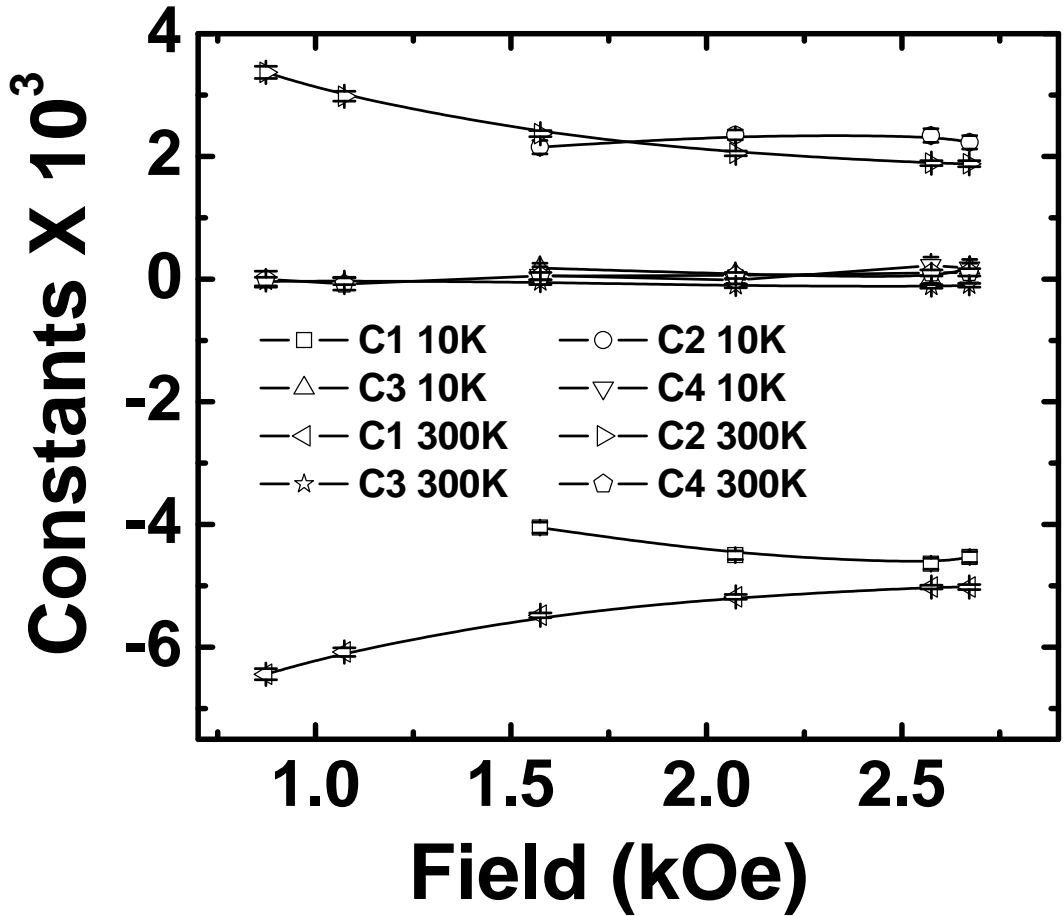}}  \caption{Field
dependence of the coefficients C1, C2, C3 and C4 obtained by fitting equation
\ref{eqnfin110} to the RMR data at 10 and 300K. In both the, cases the parameter C3
and C4 stays close to zero. The result is remarkable as it validates the fitting,
because we already know that $C_3 = C_4 = 0$ due to $\vec{I} \| \hat{n}$ which makes
$\eta = 0$. We also note that the ratio of the quadratic ($C_1$) and quadruplet
($C_2$) term remains constant at 300K in the field range of $\sim$800 Oe to
$\sim$2700 Oe. At lower temperature, however, the term appearing in 4$^{th}$ power
of $\cos\psi$ is constant where as the magnitude of the quadratic term increases
with field.} \label{const110}
\end{figure}

\begin{figure}
\centerline{\includegraphics[width=5in,angle=-90]{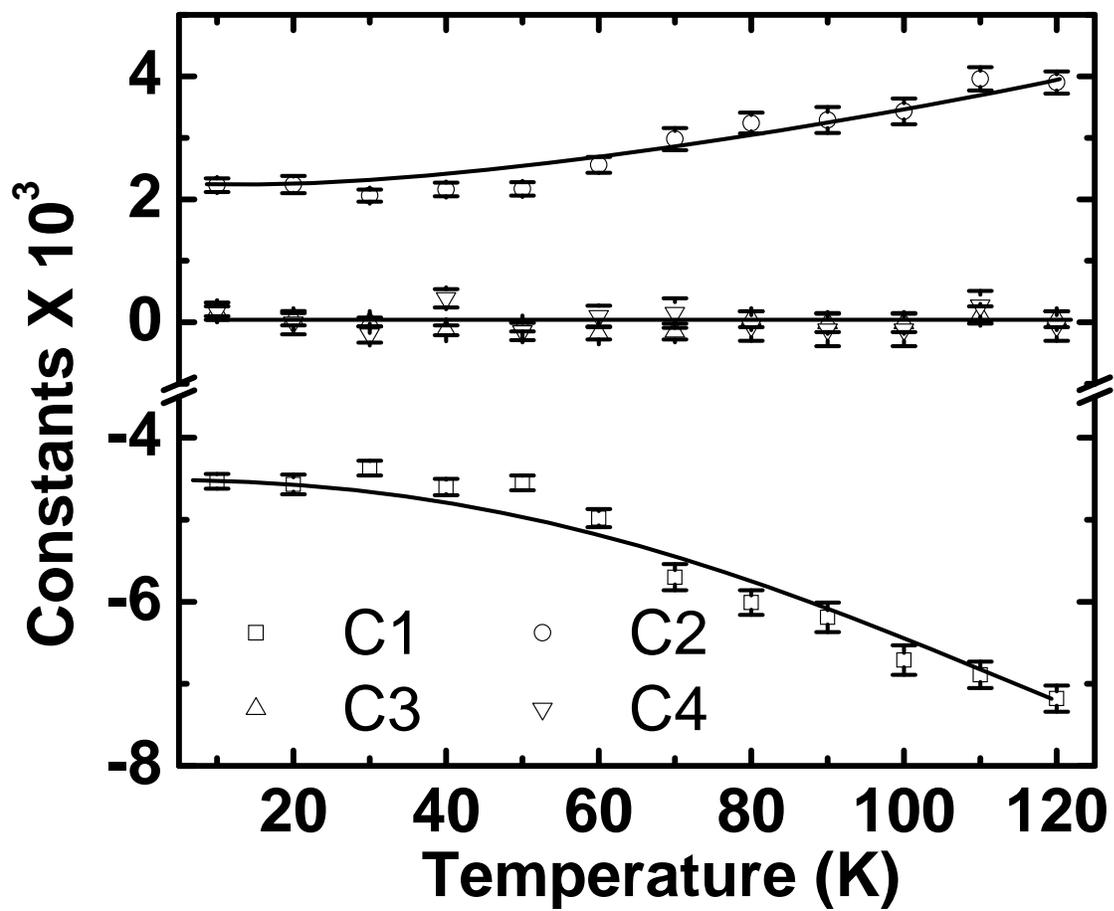}}
\caption{Temperature variation of the coefficients C1, C2, C3 and C4. The
coefficients C3 and C4 remain close to zero for all temperatures. The absolute value
of C1 and C2 increases as we go up in temperature. The solid lines are hand drawn to
indicate the most probable trend.} \label{const110temp}
\end{figure}

We have used equation \ref{eqnfin110} to fit RMR data for [110] sample at 10 and
300K, which were presented earlier in figure \ref{amr110300k} and figure
\ref{amr11010k} respectively in a compact form. The quality of fit is shown for a
representative set of data in figure \ref{fit110}.  While we see a reasonably good
fit to equation \ref{eqnfin110} down to $\simeq$ 873 Oe, the angular dependence at
still lower fields is characterized by a sharp drop in $\rho$ at $\theta > 90^0$ due
to MRPT as discussed earlier. For the 10K data, the quality of the fit is poor even
at high fields ($\approx 2000 Oe$) and it worsens when the field is reduced below
1574 Oe. Figure \ref{const110} shows the variation of various fitting parameters
with field at 10 and 300K. In both the cases, the parameter C3 and C4 stays close to
zero. This result is remarkable as it validates the fitting, because we already know
$C_3$ and $C_4$ are zero for our geometry ($\vec{I} \| \hat{n}$ \& $\eta = 0$). We
also note that the ratio of the quadratic ($C_1$) and quadruplet ($C_2$)
coefficients remains same at 300K in the field range of $\sim$ 800 to $\sim$ 2700
Oe. At lower temperature, however, the term appearing in 4$^{th}$ power of
$\cos\psi$ remains constant where as the magnitude of the quadratic term increases
with field. In figure \ref{const110temp} we have traced the variation of these
coefficients with temperature between 10 and 120K at an applied field of 2.5 kOe.
While C3 and C4 stays close to zero for all temperatures, the absolute value C1 and
C2 increases with temperature. Before we discuss the significance of these
coefficients (C's), it is pertinent to discuss the angle dependent data for the
[001] epitaxial films.

\begin{figure}
\centerline{\includegraphics[width=4in,angle=0]{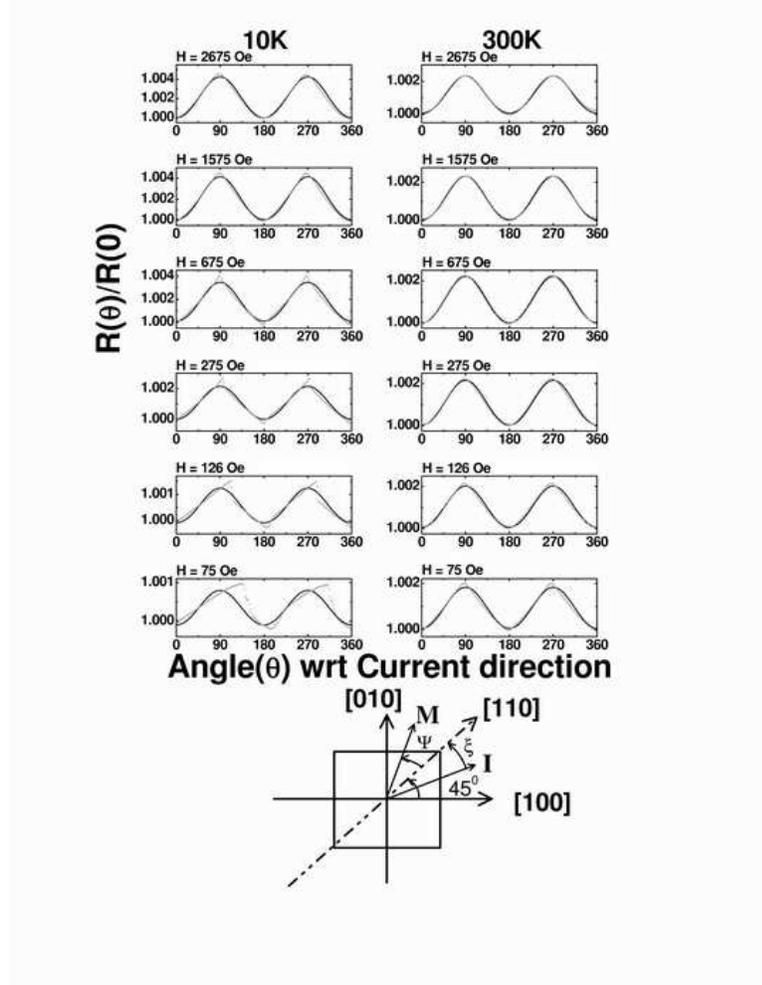}} \caption{The fit of
equation \ref{eqnfin100} to RMR data for the [001] sample at 10 and 300K.  It is
evident that at 300K the model (equation \ref{eqnfin100}) correctly describes the
behavior of the RMR down to fields as low as $\simeq$ 275 Oe. At still lower fields,
although the deviations become large, the peak and valleys of the data are correctly
reproduced. The situation, however, is quite different at 10K, here even at the
highest field deviation from the model are evident near the maxima. These deviation
becomes prominent at lower fields. A sketch of three vectors $\vec{M}$, $\vec{I}$
and the unit vector $\hat{n}$ directed along the easy axis is shown at the bottom of
the figure. In our experiment $\xi = \pi/4$ and $\psi = \theta - \pi/4$ where
$\theta$ is the angle between applied field and current direction. The current is
flowing along the hard axis and the easy axis is [110]\cite{Easy}.} \label{fit100}
\end{figure}

For the [001] film, the easy axis is along [110] where as the current is along [100]
direction. This makes the direction cosines of magnetization ($\alpha$'s) and
cosines of current ($\beta$'s) with respect to cubic axis as $\alpha_1 =
({1}/{\sqrt{2}})\left(\cos\psi - \sin\psi\right)$, $\alpha_2 =
({1}/{\sqrt{2}})\left(\cos\psi + \sin\psi\right)$ and $\alpha_3 = 0$, and $\beta_1 =
({1}/{\sqrt{2}})\left(\cos\xi + \sin\xi\right)$, $\beta_2 =
({1}/{\sqrt{2}})\left(\cos\xi - \sin\xi\right)$, $\beta_3 = 0$ and $c = 1/4$. The
final expression for the resistivity in this case is;
\begin{eqnarray} \frac{\Delta\rho}{\rho_0} - \gamma & = &
A_1\cos^2\psi + A_2\cos^4\psi + A_3\sin\psi\cos\psi\label{eqnfin100}\\
\mbox{where,}&& \;\;\;\;\;\;\nonumber
\\ A_1 & = & \frac{k_4}{3} -
k_2\left(1-2\cos^2\xi\right) -k_3\\
A_2 & = & k_3-\frac{k_4}{3}\\
A_3 & = & -2\left(k_1+k_4\right)\cos\xi\sin\xi\\
\gamma & = & \frac{k_1}{6}+k_2\left(\frac{1}{2} -
\cos^2\xi\right)+\frac{k_4}{12}\end{eqnarray} In this case $\xi = \pi/4$ and $\psi =
\theta - \pi/4$ where $\theta$ is the angle between applied field and current
direction. A sketch of three vectors $\vec{M}$, $\vec{I}$ and the unit vector
$\hat{n}$ pointing along the easy axis in a most generalized situation is given at
the bottom of figure \ref{fit100}. Since $\xi = \pi/4$, it turns out that in this
case $A_1 = -A_2$. Here we have assumed $({\Delta\rho})/({\rho_0}) - \gamma \approx
({R - R_0})/({R_0})$, where $R_0$ is the resistance when the field is aligned along
[110] the easy axis. In figure \ref{fit100}  we have shown the fit of equation
\ref{eqnfin100} to RMR data for [001] sample at 10 and 300K. It is evident that at
300K the model (equation \ref{eqnfin100}) correctly describes the behavior of the
RMR down to fields as low as $\simeq$ 275 Oe. At still lower fields although the
deviations become large, the peak and valleys of the data are correctly reproduced.
The situation, however, is quite different at 10K; here even at high fields
deviation from the model are evident near the maxima.

\begin{figure}
\centerline{\includegraphics[height=5in,angle=-90]{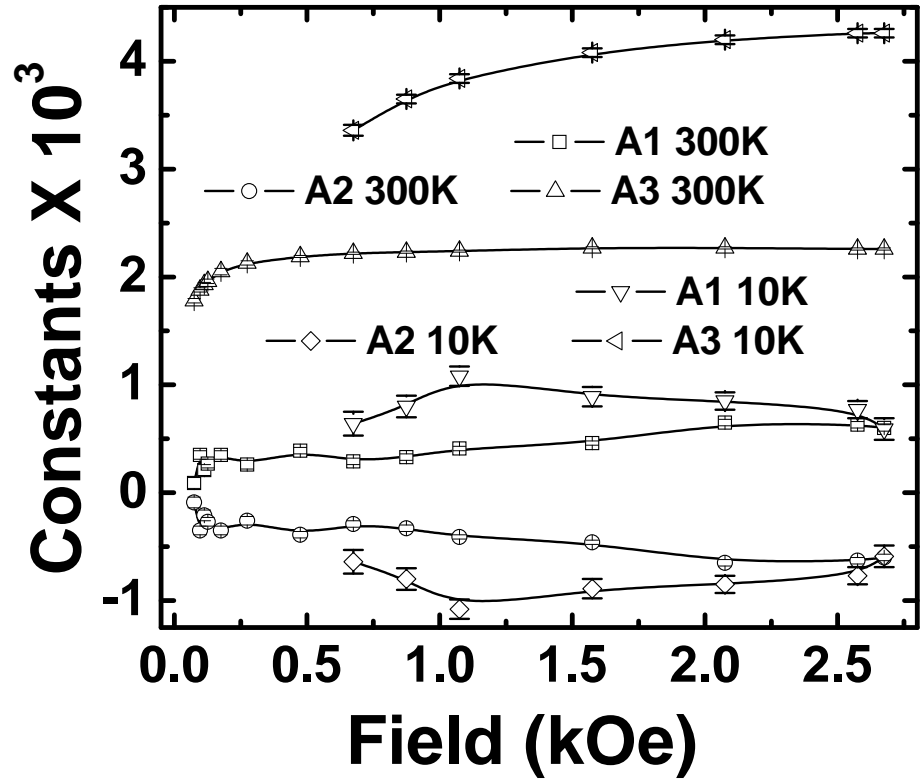}} \caption{Field
dependence of the coefficients A1, A2 and A3 obtained by fitting equation
\ref{eqnfin100} to the RMR data. Here $A_1 = -A_2$ as expected from the model under
the geometry of our measurements. The coefficient $A_3$ at both temperatures first
increases with field and then becomes constant. It should be noted that $A_3$ in
equation \ref{eqnfin100} appears as a coefficient of $\cos\psi\sin\psi$ which has
extrema at $45^0, 135^0, 225^0$ and $315^0$.} \label{const100}
\end{figure}

\begin{figure}
\centerline{\includegraphics[width=5in,angle=-90]{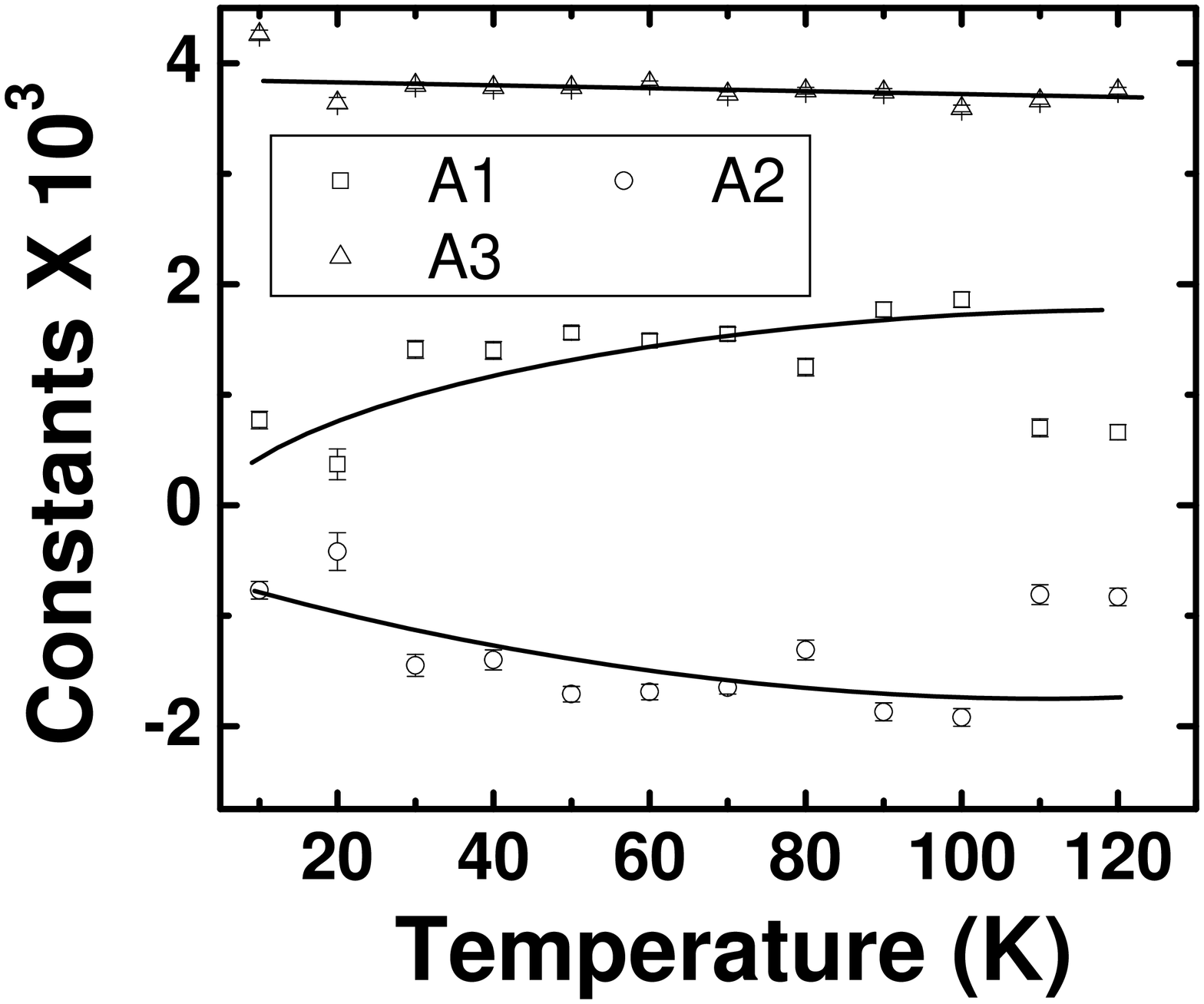}}
\caption{Temperature variation of phenomenological coefficients A1, A2 and A3. In
this case A3 remains almost constant throughout the temperature range. A1 increases
and A2 decreases as we go up in temperature. The solid lines are hand drawn to
indicate the most probable trend.} \label{const100temp}
\end{figure}

We now discuss the behavior of the coefficients $A_1, A_2$ and $A_3$ whose variation
as a function of field is shown in figure \ref{const100}. First of all, we note that
$A_1 = -A_2$ as expected from the model under the geometry of these measurements.
The coefficient $A_3$ at both temperatures increases with field and then becomes
constant at high field. Moreover, $A_3$ increases by a factor of 2 at 10K. It should
be noted that $A_3$ in equation \ref{eqnfin100} appears as a coefficient of
$\cos\psi\sin\psi$ which has extrema at $45^0, 135^0, 225^0$ and $315^0$. A higher
weightage of $A_3$ will lead to large deviations from the $\cos^2\psi$ dependence.
Figure \ref{const100temp} shows the dependence of coefficients $A_1, A_2$ and $A_3$
with temperature. We can clearly see that $A_3$ remains almost constant with
temperature. The coefficient $A_1$ increases with temperature while $A_2$ decreases
with temperature.

The data presented in figures \ref{amr110300k}, \ref{amr11010k}, \ref{amr100300K}
and \ref{amr10010K} clearly show that the RMR is both temperature and film
orientation dependent. From these data we also conclude that the [110] films have
higher anisotropy energy then the films with [001] orientation. This becomes clear
from the fact that at 300K as well as at 10K the coherent rotation of magnetization
with applied field which results in $\cos^2\psi$ dependence of RMR appears at much
higher fields for [110] films then for [001] films. Secondly, a look at the AMR
percentages calculated from these data shows that for the [110] films the absolute
value of AMR increase with temperature while it is opposite for the [001] films. We
believe that the magnetization vector of the [110] films at low temperature is
strongly pinned along the easy axis due to the large anisotropy energy. At higher
temperatures, the thermal energy $k_BT$ helps in depinning and free rotation of
$\vec{M}$ along with the external field $\vec{H}$. This leads to enhanced RMR at
higher temperatures. Of course, if the field strength is increased further, free
rotation would become possible at 10K as well. The required fields, however, may be
well beyond what has been used in these experiments. A direct support to this
argument come from the non-saturating trend of 10K AMR of the [110] film as a
function of field (figure \ref{amrper}). A very interesting picture emerges from the
value of constants so calculated. While for the [110] film the RMR is mostly
dependent on even powers of $\cos\psi$, the dependence of the RMR for the [001]
films is predominantly $\cos\psi\sin\psi$ where $\psi$ is the angle between applied
field and easy axis of film. A qualitative explanation for this observation can be
given if we refer to figure \ref{schematic} where the direction of Mn-O-Mn bond of
the [001] STO surface is shown. As we have stated earlier, the [110] direction is
the easy axis because of the degeneracy of [010] and [100] directions in zero field.
An external field applied at 45$^0$ with respect to the [110] direction lifts this
degeneracy, which perhaps is the reason why MR has a strong contribution from the
$\cos\psi\sin\psi$ term with its maximum at $\psi = 45^0$.

A discussion on the temperature and angular dependence of RMR would remain
incomplete unless we address the role of electron localization in cyclotron orbits.
The orbital magnetoresistance resulting due to electron trapping is given
as\cite{Gorkom},
\begin{equation}\left(\frac{\Delta\rho}{\rho}\right)_{OMR} =
\left(\frac{eB_\bot\tau}{m^*}\right)^2 \label{omr} \end{equation} in the limit
$\left(eB_\bot\tau/m\right)^2 \ll 1$, where $\tau$ is the relaxation time and
B$_\bot$ the component of magnetic induction perpendicular to current I. It is
expected to be negligible when carrier mean free path l is much shorter than its
cyclotron orbit (r$_c$). With the known hole density ($\approx
1.16\times10^{28}$/m$^3$), and Fermi energy (1.8eV)\cite{Stroudlsmo} of LSMO and 10K
resistivity and magnetic induction B(=H+4$\pi$M) of the [001] film, which are 0.11
m$\Omega$-cm and 7500 G for [001] films respectively we obtain $l \approx 2nm$ and
$r_c \geq 6.8 {\mu}m$. Similarly for the [110] film, the l and r$_c$ are $\approx
3nm$ and $\geq 6.6 {\mu}m$ respectively. From these numbers it can be concluded that
OMR will make negligible contribution to RMR in these films. Following
Gorkom\cite{Gorkom}, the angular dependence of OMR can be written
as;\begin{equation} \left(\frac{\Delta\rho}{\rho}\right)_{OMR} = \kappa \left(
\frac{\rho_{300}^2}{\rho^2} \right)\sin^2\theta, \label{rhoomr} \end{equation} where
$\kappa = \left(B/ne\rho_{300}\right)^2$ and $\rho_{300}$ is the resistivity at T =
300K. Using the $\rho_{10K}$ and $\rho_{300K}$ data for [001] film and equation
\ref{rhoomr} we obtain $\kappa \approx 2.3\times10^{-10}$ and
$\left(\Delta\rho/\rho\right)_{OMR} \approx 1.4\times10^{-5}\%$ which is much too
small compared to measured RMR. From this analysis it can be concluded that the
origin of angular magnetoresistance is these films is spin-orbit coupling dependent
AMR effect.

\section{Summary}
We have carried out a comparative study of the isothermal magnetoresistance of [001]
and [110]epitaxial \LSMO films as a function of the angle between current and
coplanar magnetic field at several temperatures between 10 and 300K. The magnetic
easy axis of the [001] and [110] films is along [110] and [001] directions
respectively. In view of the similar texture of these two types of films, which can
otherwise contribute to shape anisotropy, we conclude that the easy axis is
fundamentally related to the orientation of Mn-O-Mn bonds on the plane of the
substrate. The isothermal resistance $\rho_\bot$ and $\rho_\|$ for $\vec{I} \bot
\vec{H}$ and $\vec{I} \| \vec{H}$ configurations respectively of these two type of
films obeys the inequality $\rho_\bot > \rho_\|$ for all fields and temperatures.
However, the $\rho(\theta)$ shows deviation from the simple $\cos^2\theta$
dependence at low fields due to pinning of the magnetization vector $\vec{M}$ along
the easy axis. This effect manifests itself as a discontinuity in $\rho(\theta)$ at
$\theta > \pi/2$ and a concomitant hysteresis on reversing the angular scan. we
establish a magnetization reorientation phase transition in this system and extract
the H-T phase space where $\vec{M}$ remains pinned. A robust pinning of
magnetization seen in [110] films suggests strong in-plane anisotropy as compared to
the [001] films. We have carried out a full fledged analysis of the rotational
magnetoresistance of the two types of epitaxial LSMO films in the frame work of the
D\"oring theory\cite{Doring} of anisotropic magnetoresistance in metallic
ferromagnet single crystals. We note that strong deviation from the predicted
angular dependence exist in the irreversible regime of magnetization. A simple
estimation of orbital MR in these films suggest that the RMR is dominated by
spin-orbit interaction dependent anisotropic magnetoresistance.

The authors would like to thank Prof T V Ramakrishna for fruitful discussions and
Mr. Rajeev Sharma for assistance in SQUID measurements. This research has been
supported by a grant from the Board of Research in Nuclear Sciences, Government of
India.

\section*{Figure Captions}
{\bf Figure 1:} Magnetic hysteresis loops of the [110] and [001] oriented \LSMO
films at 10K. The measurements were done under the zero field cooled condition. The
direction of the applied field in these measurements was along the [001] and [110]
crystallographic directions for the [110] and [001] oriented films respectively. The
coercive field deduced from these measurements is 230 for the [110] and 90 Oe for
the [001] samples respectively.

{\bf Figure 2:} Rotational magnetoresistance R($\theta$) of the [110] film measured
at 300K for different values of the in-plane field. We observe a periodicity of
$\pi$ when the external field H is $\geq$ 300 Oe. Below 300 Oe however, a distinct
deviation from this symmetry is seen, and the peak in resistance is now shifted to
$\theta > 90^0$. One noticeable feature of the low field ($<$ 300 Oe) data is a
sudden drop in resistance once the peak value is reached. The top left inset shows
the variation of the position of first peak in the RMR data. A sketch of sample
geometry is shown in the top right hand corner of the figure.

{\bf Figure 3:} The R$(\theta)$/R(0) vs $\theta$ graphs of the [110] film at 10K for
several values of the in-plane field. Here we see deviations from a symmetric
dependence on $\theta$ at fields lower than 1100 Oe. The inset at the top left hand
corner shows the variation of peak position with field of the first peak in the RMR
data. The top right hand corner shows a sketch of the measurement geometry.

{\bf Figure 4:} Rotational magnetoresistance R($\theta$) of the [001] LSMO thin film
measured at 300K for different values of the in-plane field. We observe a
periodicity of $\pi$ for all external field H. The top left inset shows the
variation of the position of first peak in the RMR data. A sketch of sample geometry
is shown in the top right hand corner of the figure.

{\bf Figure 5:} The R$(\theta)$/R(0) vs $\theta$ graphs of the [001] film at 10K for
several values of the in-plane field. Here we see deviations from a symmetric
dependence on $\theta$ at fields lower than 500 Oe. The inset at the top left hand
corner shows the variation of peak position with field of the first peak in the RMR
data. The top right hand corner shows a sketch of the measurement geometry.

{\bf Figure 6:} Hysteresis in RMR for [110] LSMO film measured at 300K and 120 Oe .
The open circles and squares show the data when the field is varied in the forward
(0-2$\pi$) and reverse (2$\pi$-0) cycles respectively.

{\bf Figure 7:} H-T phase diagram for [110](open squares) and [001](open circles)
films. The solid lines are hand drawn to show the most probable separation line
between pinned and depinned states. These data clearly show that the pinning is much
stronger in the case of [110] film than for the [001] films. A change in y-scale
emphasizes this point.

{\bf Figure 8:} Schematic of the ionic positions on the surface of [110] and [001]
films are in the upper and lower panels respectively. For the [110] film,  the two
Mn ions along the [001] direction are bridged by an oxygen ion. Hence the [001]
direction acts as magnetic easy axis in these films. In case of [001] the [100] and
[010] direction are degenerate, hence the easy axis is along [110].

{\bf Figure 9:} Variation of the RMR percentage, defined as $\frac{R_\| -
R_\bot}{R_\bot} \times 100$, with field for both [110] and [001] LSMO at 10 and
300K. The RMR is negative for both films.

{\bf Figure 10:} Variation of RMR percentage with temperature for [110](open
squares) and [001](open circles) LSMO films between 10K and 120K. The solid lines
are hand drawn depicting the most probable trend in this temperature range. The
field applied in this case 2.5 kOe.

{\bf Figure 11:} Fit of equation \ref{eqnfin110} to RMR data of [110] film taken at
10 and 300K(left and right hand side respectively). The dots are actual data and
solid lines are fitted curve. While at 300K a reasonably good fit to equation
\ref{eqnfin110} is seen down to $\simeq$873 Oe, at still lower fields, the angular
dependence is characterized by sharp drop of $R(\theta)/R(0)$ at $\theta > 90^0$. At
these fields the torque on $\vec{M}$ exerted by the external field is not strong
enough for coherent rotation. For the 10K data, the quality of fit is poor even at
the higher fields and worsens as it is reduced below 1574 Oe. A sketch of three
vectors $\vec{M}$, $\vec{I}$ and the unit vector $\hat{n}$, and of relevant angles
is shown at the bottom of the figure. In our case $\vec{I} \| \hat{n}$ hence $\eta =
0$ and [001] is the easy axis\cite{Easy}.

{\bf Figure 12:} Field dependence of the coefficients C1, C2, C3 and C4 obtained by
fitting equation \ref{eqnfin110} to the RMR data at 10 and 300K. In both the, cases
the parameter C3 and C4 stays close to zero. The result is remarkable as it
validates the fitting, because we already know that $C_3 = C_4 = 0$ due to $\vec{I}
\| \hat{n}$ which makes $\eta = 0$. We also note that the ratio of the quadratic
($C_1$) and quadruplet ($C_2$) term remains constant at 300K in the field range of
$\sim$800 Oe to $\sim$2700 Oe. At lower temperature, however, the term appearing in
4$^{th}$ power of $\cos\psi$ is constant where as the magnitude of the quadratic
term increases with field.

{\bf Figure 13:} Temperature variation of the coefficients C1, C2, C3 and C4. The
coefficients C3 and C4 remain close to zero for all temperatures. The absolute value
of C1 and C2 increases as we go up in temperature. The solid lines are hand drawn to
indicate the most probable trend.

{\bf Figure 14:} The fit of equation \ref{eqnfin100} to RMR data for the [001]
sample at 10 and 300K.  It is evident that at 300K the model (equation
\ref{eqnfin100}) correctly describes the behavior of the RMR down to fields as low
as $\simeq$ 275 Oe. At still lower fields, although the deviations become large, the
peak and valleys of the data are correctly reproduced. The situation, however, is
quite different at 10K, here even at the highest field deviation from the model are
evident near the maxima. These deviation becomes prominent at lower fields. A sketch
of three vectors $\vec{M}$, $\vec{I}$ and the unit vector $\hat{n}$ directed along
the easy axis is shown at the bottom of the figure. In our experiment $\xi = \pi/4$
and $\psi = \theta - \pi/4$ where $\theta$ is the angle between applied field and
current direction. The current is flowing along the hard axis and the easy axis is
[110]\cite{Easy}.

{\bf Figure 15:} Field dependence of the coefficients A1, A2 and A3 obtained by
fitting equation \ref{eqnfin100} to the RMR data. Here $A_1 = -A_2$ as expected from
the model under the geometry of our measurements. The coefficient $A_3$ at both
temperatures first increases with field and then becomes constant. It should be
noted that $A_3$ in equation \ref{eqnfin100} appears as a coefficient of
$\cos\psi\sin\psi$ which has extrema at $45^0, 135^0, 225^0$ and $315^0$.

{\bf Figure 16:} Temperature variation of phenomenological coefficients A1, A2 and
A3. In this case A3 remains almost constant throughout the temperature range. A1
increases and A2 decreases as we go up in temperature. The solid lines are hand
drawn to indicate the most probable trend.

\begin{thebibliography}{}
\bibitem{Robert}
Robert C. O'Handley Modern Magnetic Materials Principles and Applications, 2000,
(John Wiley and Sons, Inc., New York) pp. 579-584

\bibitem{McGuire}
T. R. McGuire and R. I. Potter, IEEE Trans. Mag. {\bf 11}, 1018 (1975).

\bibitem{Fert}
O. Jaoul, I. A. Campbell and A. Fert, J. Magn. Magn. Mater. {\bf 5}, 23 (1977).

\bibitem{Malozemoff}
A. P. Malozemoff, Phys. Rev. B {\bf 32}, 6080 (1985).

\bibitem{Ecksteinapl1996}
J. N. Eckstein, I. Bozovic, J. O'Donnell, M. Onellion and M. S. Rzchowski, Appl.
Phys. Lett. {\bf 69}, 1312 (1996).

\bibitem{Ziese}
M. Ziese and S. P. Sena, J. Phys: Condens. Matter {\bf 10}, 2727 (1998).

\bibitem{Ziesesing}
M. Ziese, Phys. Rev. B {\bf 62}, 1044 (2000).

\bibitem{Infante}
I. C. Infante, D. Hrabovsk\'y, V. Laukhin, F. S\'anchez and J. Fontcuberta, J. Appl.
Phys. {\bf 99}, 08C503 (2006).

\bibitem{Hong}
X. Hong, J. -B. Yau, J. D. Hoffman, C. H. Ahn, Y. Bason and L. Klein, Phys. Rev. B
{\bf 74}, 174406 (2006).

\bibitem{Hamaya}
K. Hamaya, R. Moriya, A. Oiwa, T. Taniyama, Y. Kitamoto Y and H. Munekata, IEEE
Trans. Mag. {\bf 39}, 2785 (2003).

\bibitem{Gns}
G. Singh and R. C. Budhani, {\it unpublished}.

\bibitem{Prinzjap}
E. Dan Dahlberg, Kevin Riggs and G. A. Prinz, J. Appl. Phys. {\bf 63}, 4270 (1988).

\bibitem{Prinzprb}
K. T. Riggs, E. Dan Dahlberg and G. A. Prinz, Phys. Rev. B {\bf 41}, 7088 (1990).

\bibitem{Gorkom}
R. P. van Gorkom, J. Caro, T. M. Klapwijk and S. Radelaar, Phys. Rev. B {\bf 63},
134432 (2001).

\bibitem{Zener}
Clarence Zener, Phys. Rev. {\bf 82}, 403 (1951).

\bibitem{Anderson}
P. W. Anderson and H. Hasegawa, Phys. Rev. {\bf 100}, 675 (1955).

\bibitem{Gennes}
P. -G. de Gennes, Phys. Rev. {\bf 118}, 141 (1960).

\bibitem{Tyagi}
S. D. Tyagi, S. E. Lofland, M. Dominguez, S. M. Bhagat, C. Kwon, M. C. Robson, R.
Ramesh and T. Venkatesan, Appl. Phys. Lett. {\bf 68}, 2893 (1996).

\bibitem{Snyder}
G. Jeffrey Snyder, M. R. Beasley, T. H. Geballe, Ron Hiskes and Steve DiCarolis,
Appl. Phys. Lett. {\bf 69}, 4254 (1996).

\bibitem{Doring}
W. D\"oring, Ann. Phys. {\bf 424}, 259 (1938).

\bibitem{Senapati}
K. Senapati and R. C. Budhani, Phys. Rev. B {\bf 71}, 224507
(2005).

\bibitem{Patnaik}
S. Patnaik, Kanwaljeet Singh and R. C. Budhani, Rev. Sci. Instr.
{\bf 17}, 1494 (1999).

\bibitem{Easy}
The easy axis of both [110] and [001] oriented films were determined by measuring M
vs H, by applying fields at different angles with respect to a known
crystallographic direction in the plane of the film.

\bibitem{Suzuki}
L. M. Berndt, Vincent Balbarin and Y Suzuki, Appl. Phys. Lett. {\bf 77}, 2903
(2000).

\bibitem{Fisher}
David Mukamel, Michael E. Fisher and Eytan Domany, Phys. Rev. Lett. {\bf 37}, 565
(1976).

\bibitem{Tsui}
F. Tsui, M. C. Smoak, T. K. Nath and C. B. Eom, Appl. Phys. Lett. {\bf 76}, 2421
(2000).

\bibitem{Lecoeur}
P. Lecoeur, P. L. Trouilloud, Gang Xiao, A. Gupta, G. Q. Gong and X. W. Li, J. Appl.
Phys. {\bf 82}, 3934 (1997).

\bibitem{Muduli}
P K Muduli, S K Bose and R C Budhani, J. Phys.: Condens. Matter {\bf 19}, 226204
(2007)

\bibitem{Malozemoffprb}
A. P. Malozemoff, Phys. Rev. B {\bf 34}, 1853 (1986).

\bibitem{Herranz}
G. Herranz, F. S\'anchez, M. V. Garc\'ia-Cuenca, C. Ferrater, M. Varela, B.
Mart\'inez and J. Fontcuberta, J. Magn. Magn. Mater. {\bf 272-276}, 517 (2004).

\bibitem{Bibes}
M. Bibes, O. Gorbenko, B. Mart\'inez, A. Kaul and J. Fontcuberta, J Magn. Magn.
Mater. {\bf 211}, 47 (2000).

\bibitem{Stroudlsmo}
B. Nadgorny, I. I. Mazin, M. Osofsky, R. J. Soulen Jr., P. Broussard, R. M. Stroud,
D. J. Singh, V. G. Harris, A. Arsenov and Ya Mukovskii, Phys. Rev. B {\bf 63},
184433 (2001).

\end{thebibliography}
\end{document}